\documentclass{article}

\usepackage[preprint]{neurips_2025}

\usepackage[utf8]{inputenc} 
\usepackage[T1]{fontenc}    
\usepackage{hyperref}       
\usepackage{url}            
\usepackage{booktabs}       
\usepackage{amsmath}        
\usepackage{amsfonts}       
\usepackage{nicefrac}       
\usepackage{microtype}      
\usepackage{xcolor}         
\usepackage{comment}
\usepackage{etoolbox}
\usepackage{soul}
\usepackage{amssymb}
\usepackage{graphicx}
\usepackage{makecell}
\usepackage{amsthm}
\usepackage{subcaption}
\usepackage{ifthen}

\newcommand{\tightinlinebox}[1]{%
  \setlength{\fboxsep}{1pt}%
  \fcolorbox{black}{white}{#1}%
}

\newcommand{\maybeCite}[1]{%
  \ifthenelse{\equal{#1}{}}%
    {}%
    {\citep{#1}}%
}

\newcommand{\ours}{\textsc{MagiCodec}}

\newtheorem{proposition}{Proposition}

\title{MagiCodec: Simple Masked Gaussian-Injected Codec for High-Fidelity Reconstruction and Generation}

\author{
    Yakun Song\textsuperscript{1,2}\thanks{Equal contribution.}
    \And Jiawei Chen\textsuperscript{2}\footnotemark[1]
    \And Xiaobin Zhuang\textsuperscript{2}\footnotemark[1]
    \And Chenpeng Du\textsuperscript{2}
    \AND Ziyang Ma\textsuperscript{1,2}
    \And Jian Wu\textsuperscript{2}
    \And Jian Cong\textsuperscript{2}
    \And Dongya Jia\textsuperscript{2}
    \AND Zhuo Chen\textsuperscript{2}
    \And Yuping Wang\textsuperscript{2}
    \And Yuxuan Wang\textsuperscript{2}
    \And Xie Chen\textsuperscript{1}
    \AND \\ \textsuperscript{1}Shanghai Jiao Tong University
    \And \\ \textsuperscript{2}Bytedance Inc.
}

\begin{document}

\maketitle

\begin{abstract}
Neural audio codecs have made significant strides in efficiently mapping raw audio waveforms into discrete token representations, which are foundational for contemporary audio generative models. However, most existing codecs are optimized primarily for reconstruction quality, often at the expense of the downstream modelability of the encoded tokens. Motivated by the need to overcome this bottleneck, we introduce \textbf{MagiCodec}, a novel single-layer, streaming Transformer-based audio codec. MagiCodec is designed with a multistage training pipeline that incorporates Gaussian noise injection and latent regularization, explicitly targeting the enhancement of semantic expressiveness in the generated codes while preserving high reconstruction fidelity. We analytically derive the effect of noise injection in the frequency domain, demonstrating its efficacy in attenuating high-frequency components and fostering robust tokenization. Extensive experimental evaluations show that MagiCodec surpasses state-of-the-art codecs in both reconstruction quality and downstream tasks. Notably, the tokens produced by MagiCodec exhibit Zipf-like distributions, as observed in natural languages, thereby improving compatibility with language-model-based generative architectures. The code and pre-trained models are available at \url{https://github.com/Ereboas/MagiCodec}.
\end{abstract}

\section{Introduction}

Recently, large language models (LLMs) have made transformative progress in natural language processing\maybeCite{achiam2023gpt, liu2024deepseek, yang2024qwen2} and audio generation\maybeCite{anastassiou2024seed, borsos2023audiolm}, exhibiting a strong capability to model long sequences of discrete tokens.
Recent studies have widely adopted audio codec models, such as SoundStream\maybeCite{zeghidour2021soundstream} and EnCodec\maybeCite{defossez2022high}, as audio tokenizers within audio language modeling frameworks. However, these methods continue to focus on fidelity and computational efficiency in reconstruction as their primary objectives, and often overlook the semantic modelability of discrete representations.

As research in representation learning progresses, many researchers have increasingly recognized the \textit{optimization dilemma} between generative capacity and reconstruction quality. Specifically, improving reconstruction quality can compromise generation performance and require larger models and more training resources. Conversely, limiting reconstruction capabilities can reduce the upper limit of generation quality\maybeCite{yao2025reconstruction}. Therefore, overemphasis on reconstruction objectives tends to substantially complicate the training of generative models.

To enhance generative performance, several studies have introduced explicit semantic supervision to strengthen the encoding of low-frequency semantic content.
For example, SemantiCodec\maybeCite{liu2024semanticodec} employs a dual-encoder architecture, combining high-level semantic features extracted by a self-supervised, pretrained AudioMAE\maybeCite{huang2022masked} with an acoustic encoder that captures fine details, and utilizes a diffusion-based decoder to achieve high-quality reconstruction at ultra-low bitrates. X-Codec\maybeCite{ye2025codec} integrates self-supervised semantic representations directly into the quantization process, thereby enhancing the generative capability of audio language models.
Although these strategies yield improved semantic retention, they also result in the loss of high-frequency texture details and the introduction of minor artifacts. In addition, they depend on external models, such as diffusion models\maybeCite{anastassiou2024seed}, and thus fail to provide an efficient or fundamental solution. The question of how to achieve high-fidelity reconstruction and improved modelability of discrete codes through intrinsic frequency-domain constraints or regularization mechanisms, without resorting to additional annotation or complex pretraining, remains a central issue in contemporary audio codec research.

Another major challenge for neural audio codecs arises from their underlying architecture. A typical neural codec can be divided into three components: an encoder that projects the raw waveform into a continuous latent vector space; a vector quantization (VQ) module that discretizes the latent vectors into tokens drawn from a finite codebook\maybeCite{van2017neural}; and a decoder that reconstructs the audio signal from these discrete tokens\maybeCite{wu2024ts3}. The VQ module is the central element in achieving discretization. During training, straight-through gradient estimation is typically employed; although this simplifies backpropagation, it also amplifies the error introduced by quantization. Whether using vanilla vector quantization or residual vector quantization (RVQ)\maybeCite{zeghidour2021soundstream, defossez2022high}, training commonly suffers from codebook collapse, in which many codebook entries go underutilized or only a small subset of vectors is frequently activated. This phenomenon leads to poor coverage of the encoding space and reduced token diversity, thereby undermining the expressive capacity and efficiency of downstream generative models.

In response to these challenges, we propose MagiCodec, a simple \textbf{\tightinlinebox{Ma}sked \tightinlinebox{G}aussian \tightinlinebox{I}njected \tightinlinebox{Codec}} for high-fidelity reconstruction and generation. Largely inspired by TS3Codec\maybeCite{wu2024ts3}, MagiCodec also adopts an efficient, single-layer streaming codec built on a Transformer backbone. Through a multi-stage training procedure with clearly articulated motivations, it implicitly suppresses high-frequency noise and strengthens low-frequency structure modeling, thereby achieving joint optimization of reconstruction quality and downstream generative performance. Our core idea is to dispense with external labels and rely solely on intrinsic Gaussian noise injection, so that the codec learns to allocate modeling capacity appropriately across different frequency bands, achieving both high-fidelity reconstruction and efficient, modelable discrete codes at constrained bitrates. In addition, we decompose traditional codec training into three phases: autoencoder, vector quantization, and vocoder. Training in the first two stages involves only the generator, thereby preventing audio phase information from being incorporated into the intermediate representations. This staged training approach effectively avoids the issue of codebook collapse and maximizes codebook utilization. Our main contributions can be summarized as follows:

\begin{itemize}
\item \textbf{Theoretical contributions.} We analytically derive the frequency-domain effect of Gaussian noise injection, showing that it is equivalent to mixing the original signal with a low-pass–filtered version, thereby implicitly regularizing high-frequency components.

\item \textbf{Methodological contributions.} We design a multi-stage training framework that effectively mitigates codebook collapse and enhances overall token performance, incorporating Gaussian noise injection and latent regularization. These techniques require no external models or labels, yet encourage the codec to learn low-frequency semantic representations and avoid overfitting to high-frequency noise.

\item \textbf{Experimental contributions.} We conduct comprehensive evaluations, demonstrating that MagiCodec achieves state-of-the-art reconstruction quality across multiple bitrates and metrics. In downstream tasks such as text-to-speech, automatic speech recognition, and semantic information extraction, MagiCodec also significantly outperforms baseline methods, confirming its strong modelability. Analysis of the code distribution further reveals its close adherence to the Zipf distribution observed in natural language, which facilitates downstream model training. 
\end{itemize}

\section{Related Work}
Neural audio codecs aim to encode continuous audio signals into discrete latent representations, which is a form of discrete audio tokenization. In recent years, neural audio codecs have become a research focus to achieve high-quality audio reconstruction at low bitrates\maybeCite{kumar2023high,ai2024apcodec}. These methods leverage vector quantization to learn compact codebooks that effectively compress audio information while preserving essential details. To further improve reconstruction quality, various techniques have been proposed. PromptCodec\maybeCite{pan2024promptcodec} introduces additional input prompts to enrich the latent space representation, improving the model’s adaptability to complex audio content. Moreover, DAC\maybeCite{kumar2023high} combines quantizer dropout with multi-scale STFT discriminators, effectively enhancing spectral recovery accuracy and improving the naturalness and clarity of reconstructed audio. FreeCodec\maybeCite{zheng2024freecodec} enables efficient compression and high-quality reconstruction by decoupling inherent properties of speech (such as tone, rhythm and content). APCodec\maybeCite{ai2024apcodec} and Apcodec+\maybeCite{du2024apcodec+} integrate a staged training strategy to significantly enhance the representational power of both encoder and decoder, leading to superior reconstruction performance.
However, an excessive focus on reconstruction quality often increases the complexity of the latent space, which in turn imposes a heavier training burden on the generation model and reduces generation efficiency\maybeCite{yao2025reconstruction}. This inherent tension between reconstruction and generation remains a core challenge in the design of discrete audio tokenization.

To address this limitation, recent works like SemanticCodec\maybeCite{liu2024semanticodec}, X-Codec\maybeCite{ye2025codec} and VQGAN-LC\maybeCite{zhu2024scaling} have introduced explicit supervision from pretrained models to enhance semantic retention. While these approaches improve semantic representation, they may lead to the loss of high-frequency details and increase reliance on external resources.

In contrast, MagiCodec leverages frequency-domain regularization and Gaussian noise injection to significantly enhance token semantic expressiveness without requiring external supervision, while maintaining high-fidelity reconstruction quality.

\section{MagiCodec}
In this section, we introduce MagiCodec, a simple yet efficient single-layer streaming codec that delivers both high-fidelity reconstruction and strong downstream task performance.

\begin{figure}[t]
    \centering
    \includegraphics[width=0.7\linewidth]{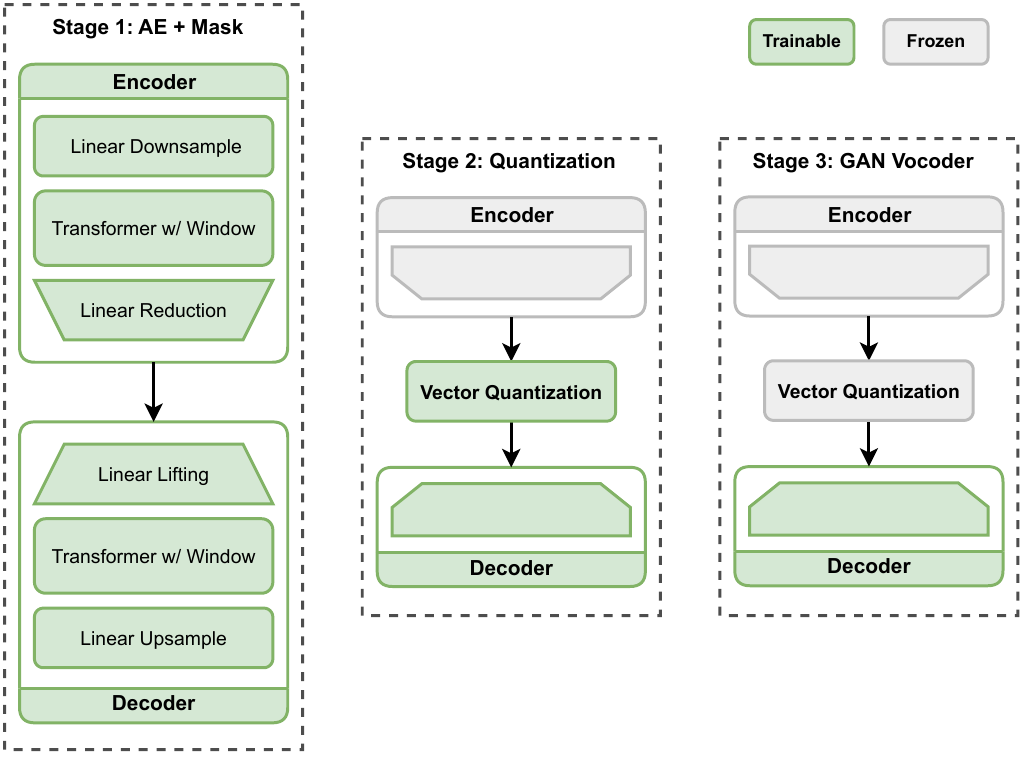}
    \caption{The pipeline of the proposed \ours.} 
    \label{fig:pipeline}
\end{figure}

\subsection{Model Architecture}

In Figure~\ref{fig:pipeline}, we present a high-level overview of MagiCodec’s end-to-end workflow.
As in traditional audio codec frameworks\maybeCite{}, the raw waveform is first downsampled and mapped into a low-dimensional latent space by the encoder, which is then discretized into tokens by the quantizer. Finally, the decoder reconstructs the waveform from these tokens.

The encoder consists of a linear downsampling module, a windowed Transformer, and a linear reduction layer. For the 16kHz model, all input audio is sampled at 16kHz. 
The encoder first applies a two-layer linear network (the linear downsampling module) to downsample the input waveform $\mathbf{x}\in\mathbb{R}^T$ by a factor $r\in\{160,320,640\}$, corresponding to token rates $T_r\in\{100,50,25\}\,\mathrm{Hz}$, and projects it into a hidden space of dimension $H=4096$, yielding the Transformer input $\mathbf{X}\in\mathbb{R}^{T_r\times H}$.
This frame sequence is then fed into a Transformer~\maybeCite{vaswani2017attention} with a sliding window of size 32, in which each token attends only to itself and its left context to enforce strict streaming inference.
A single linear projection (the linear reduction layer) maps the dimension from $H$ down to the VQ codebook embedding size $D=16$, producing $\mathbf{Z}_e\in\mathbb{R}^{T_r\times D}$.

After the encoder module, the single quantizer uses a codebook of size $K=131072$ to quantize $\mathbf{Z}_e$ into discrete tokens. During training, gradients are passed through a straight-through estimator (STE)~\maybeCite{bengio2013estimating}. At inference time, each feature is quantized to its nearest neighbor in the codebook, yielding discrete tokens.

In the decoding stage, discrete tokens are converted back to embeddings $\mathbf{Z}_q$ via codebook lookup, then passed through the decoder, which mirrors the encoder’s architecture. A single linear lifting layer restores the dimension to $H$, followed by a Transformer with a left-context window of 32 and a right-context window of 2 to enhance reconstruction quality while preserving streaming properties. Finally, a linear upsampling layer reconstructs the waveform $\hat{\mathbf{x}}\in\mathbb{R}^T$ at the original sampling rate. To further improve perceptual fidelity, MagiCodec incorporates GAN-based optimization during certain training phases. We elaborate on this in Section~\ref{sec:stages}.

\subsection{Gaussian Noise Injection}

\subsubsection{Motivation} 
Conventional audio codec research has predominantly emphasized reconstruction quality, often neglecting the dimension of generative performance\maybeCite{}. 
However, in downstream generative tasks, reconstruction fidelity, compression efficiency, and modelability are all indispensable: inaccurate reconstruction directly constrains the fidelity of generated signals; suboptimal compression efficiency not only slows down generation but also substantially increases computational and storage costs; and insufficient modelability typically forces the use of larger, more expensive, and more complex language or other generative model backbones, further exacerbating system compute demands and limiting overall performance\maybeCite{skorokhodov2025improving}.

Recent works have integrated semantic labels and other supervisory signals into VQ training to enhance the downstream modelability of tokens\maybeCite{liu2024semanticodec, zhang2023speechtokenizer, huang2023repcodec, kyutai2024moshi}. Although this semantic supervision helps the model capture low-frequency structural information, it typically requires additional neural networks, and models often sacrifice high-frequency texture details to minimize semantic loss, resulting in reconstruction artifacts\maybeCite{ye2025codec}.

Neural networks inherently exhibit a \textit{spectral bias}\maybeCite{rahaman2019spectral}, preferentially learning low-frequency structures first; high-frequency local oscillations are harder to fit and more prone to overfitting noise. 
Excessively preserving high-frequency content both wastes bits and increases the difficulty of model fitting.

We hypothesize that preserving excessive random high-frequency components degrades both the perceptual quality of the latent representation and its modelability in downstream generative tasks. To address this, we introduce Gaussian noise injection, a general-purpose method that requires no additional supervision. Our approach is exceptionally simple to implement and, from a Fourier-analytic perspective, can be shown to impose an exponentially decaying regularization on high-frequency components. As a result, it maintains high reconstruction fidelity while markedly enhancing the codec’s performance on downstream tasks.

\subsubsection{Method} 
For the input frame ${\mathbf{X}}$, we independently sample per-frame masks $m_t \sim \mathrm{Bernoulli}(p)$.
If \(m_t = 1\), the original frame is fully replaced by i.i.d. Gaussian noise 
\(\epsilon_t \sim \mathcal{N}(\mathbf{0}, \sigma^2 \mathbf{I})\); otherwise, it remains unchanged, yielding the noise-injected sequence \(\tilde{\mathbf{X}}\). 
Training with random additive noise has been shown to be equivalent to incorporating a Tikhonov regularization term into the loss function~\maybeCite{bishop1995}, and subsequent studies have further interpreted this as an explicit high-frequency regularizer, thereby encouraging the model to focus on semantically relevant low-frequency structures~\maybeCite{camuto2020explicit}.
By using replacement noise instead of additive noise, we entirely remove local time-domain information in masked frames, forcing the model to rely on longer-range context for reconstruction and overall semantics. Leveraging longer contextual dependencies enables language models to learn smoother, low-frequency–dominated latent dynamics, thereby reducing the receptive field required by downstream language models to achieve comparable semantic coverage~\maybeCite{}.
We state a concise proposition that, in the Fourier domain, elucidates the high-frequency attenuation induced by Gaussian noise injection.

\begin{proposition}
\label{prop1}
For any Fourier-transformable network mapping \(f\), applying Gaussian noise injection to the input $\mathbf{x}$, we have
\[
\mathbb{E}\bigl[f(\tilde{\mathbf{x}})\bigr]
=\Bigl[(1-p)+p\,e^{-\tfrac{1}{2}\sigma^2\|\boldsymbol{\omega}\|^2}\Bigr]\widehat{f}(\boldsymbol{\omega})\,. 
\]
\end{proposition}
Here, \(\widehat{f}(\boldsymbol{\omega})\) denotes the Fourier coefficient of \(f\) at frequency \(\boldsymbol{\omega}\).
See Appendix~\ref{appendix:prop1} for the full formulations and derivations.

Consequently, high-frequency components are explicitly attenuated, whereas low-frequency structures remain virtually unaffected. The Experiments~\ref{sec:exp} further validates the effectiveness of our proposed method.

\subsection{Training Stages}
\label{sec:stages}

In single-stage end-to-end VQ training, an unpretrained encoder coupled with a randomly initialized codebook often maps a large fraction of inputs to nearly identical embeddings, causing VQ collapse and consequently ineffective quantization, distorted generation, or even complete training stagnation. Prior work\maybeCite{zhao2024representation} categorizes this failure into token collapse and embedding collapse, attributing the root cause to the synchronous cold start of both the encoder and the codebook. To mitigate this, we employ a three-stage training strategy combined with latent-variable regularization, which in our experiments yields significantly more stable codebook learning and improved reconstruction and generative metrics. 
Specifically, our training stages are as follows:

\paragraph{Stage 1: Autoencoder}
We first train only the encoder-decoder with no quantization applied so that it learns stable representations. This warm-up provides a strong initialization for the subsequent quantization stage and prevents the synchronous oscillations that can arise when the codebook and encoder receive simultaneous gradient updates in early training\maybeCite{}.
Additionally, we incorporate a latent-space regularization loss $L_{\mathrm{norm}} = \|\mathbf{Z}_e\|_2^2$ into the training objective. This latent regularization can prevent unstable, unconstrained latent vectors from causing training collapse in the early stages.
It can be viewed as a simplified form of KL regularization, encouraging the latent space to be more compact and continuous, which is advantageous for vector quantization.

\paragraph{Stage 2: Quantizer}
During this phase, we freeze the encoder and exclusively optimize the vector quantizer and the decoder. Fed by the high-quality continuous latent representations, the quantizer can learn more robustly, mitigating codebook collapse caused by early-stage oscillations. The codebook is optimized using an L1 loss, computed between the features prior to and following quantization, with a stop-gradient operation applied as described in\maybeCite{van2017neural}. Consistent with the methodology adopted in SimVQ\maybeCite{zhu2024addressing}, we reparameterize the code vectors through a linear transformation layer based on a learnable latent basis. Additionally, to prevent the encoder outputs from attaining excessively large magnitudes, a commitment loss with a weighting factor of 0.25 is introduced.

\paragraph{Stage 3: Vocoder}
At this stage, the parameters of both the encoder and the vector quantizer are frozen, and only the decoder is updated during training. A multi-scale Mel-spectrogram reconstruction loss is employed, defined as the L1 distance between the predicted and reference spectrograms across multiple frequency resolutions. The mel-spectrogram is widely acknowledged as a reliable proxy for perceptual audio quality. Furthermore, to enhance the perceptual realism of the reconstructed audio, we adopt the adversarial training strategy introduced in BigCodec\maybeCite{xin2024bigcodec}, incorporating two distinct types of discriminators. The first is the Multi-Period Discriminator (MPD) from HiFi-GAN\maybeCite{kong2020hifi}, which is designed to capture diverse periodic structures in speech waveforms. The second is the Multi-Scale Short-Time Fourier Transform (MS-STFT) Discriminator, as implemented in EnCodec\maybeCite{defossez2022high}, which captures spectral features at multiple time-frequency resolutions.

\paragraph{Training objectives}
The training losses for the generator of MagiCodec include mel-spectrogram reconstruction loss $\mathcal{L}_{mel}$, quantizer loss $\mathcal{L}_{q}$, latent regularization loss $\mathcal{L}_{e}$, adversarial loss $\mathcal{L}_{adv}$, and feature matching loss $\mathcal{L}_{feat}$. Specifically, our training loss is formulated as follows:
For mel-spectrogram reconstruction loss, we compute the $L_1$ distance between the predicted and reference mel-spectrograms at multiple frequency resolutions.
For GAN loss, we calculate the $L_2$-norm as the adversarial loss over the logits of the discriminators and use the $L_1$-norm to calculate the feature matching loss.
For VQ loss, we follow the classic VQ training scheme, using an $L_1$ codebook loss and an $L_1$ commitment loss.

Thus, the total loss for the $i$-th training stage of MagiCodec, $\mathcal{L}^i$, is given by:
\[
\mathcal{L}^1 = \lambda^1_{mel}\mathcal{L}^1_{mel} + \lambda^1_{e}\mathcal{L}^1_{e}, \;
\mathcal{L}^2 = \lambda^2_{mel}\mathcal{L}^2_{mel} + \lambda^2_{q}\mathcal{L}^2_{q}, \; \text{and} \;
\mathcal{L}^3 = \lambda^3_{mel}\mathcal{L}^3_{mel} + \lambda^3_{adv}\mathcal{L}^3_{adv} + \lambda^3_{feat}\mathcal{L}^3_{feat}.
\]

\section{Experiments}
\label{sec:exp}

To comprehensively assess MagiCodec’s reconstruction quality and downstream generative modelability, we carried out extensive experiments.

\subsection{Experimental Setup}
\subsubsection{Datasets}
We train our codec models on the Libri-light corpus\maybeCite{kahn2020libri}, which contains approximately 60{,}000 hours of unlabelled English speech sampled at 16~kHz. We evaluate reconstruction fidelity on the LibriSpeech test-clean subset, comprising 2,620 utterances from 40 speakers.

\subsubsection{Evaluation Metrics}
We employ a range of evaluation metrics spanning several dimensions. For computational efficiency, we consider model parameter count (nParams), bitrate, frame rate, token rate, streaming capability, and the number of codebook layers. Speech intelligibility is assessed using Short-Time Objective Intelligibility (STOI), Word Error Rate (WER), and Phone Error Rate (PER). Distortion and perceptual audio quality are evaluated with metrics such as Perceptual Evaluation of Speech Quality (PESQ), Virtual Speech Quality Objective Listener (ViSQOL), and UTokyo-SaruLab MOS (UTMOS). To assess speaker similarity, we report SPK-SIM. Detailed definitions and calculation procedures for each metric are provided in Appendix~\ref{appendix:metrics}.

\subsubsection{Baselines}

We selected a series of state-of-the-art codecs as baselines. To ensure a fair comparison, we employed the official pretrained weights for EnCodec\footnote{\url{https://huggingface.co/facebook/encodec_24khz}}\maybeCite{defossez2022high}, Mimi\footnote{\url{https://huggingface.co/kyutai/mimi}}\maybeCite{kyutai2024moshi}, DAC\footnote{\url{https://huggingface.co/descript/dac_16khz}}\maybeCite{kumar2023high}, Vocos\footnote{\url{https://huggingface.co/charactr/vocos-mel-24khz}}\maybeCite{siuzdak2023vocos}, SNAC\footnote{\url{https://github.com/hubertsiuzdak/snac}}\maybeCite{siuzdak2024snac}, WavTokenizer\footnote{\url{https://github.com/jishengpeng/WavTokenizer}}\maybeCite{ji2024wavtokenizer}, SpeechTokenizer\footnote{\url{https://github.com/ZhangXInFD/SpeechTokenizer}}\maybeCite{zhang2023speechtokenizer}, and SemantiCodec\footnote{\url{https://github.com/haoheliu/SemantiCodec-inference}}\maybeCite{liu2024semanticodec}.
Given that TS3Codec serves as our primary baseline and no official pretrained model is available, we directly extracted the experimental results for both TS3Codec and BigCodec-S from the TS3Codec paper\maybeCite{wu2024ts3}. Here, BigCodec-S refers to the streaming variant of BigCodec\maybeCite{xin2024bigcodec} implemented by the TS3Codec authors using the official BigCodec code. We adopt the identical training and evaluation datasets used by TS3Codec and BigCodec-S to ensure a fair comparison (see Table \ref{tab:computation} for more details for various codec models).

\begin{table}[t]
  \centering
  \caption{Computation efficiency comparison of various codec models.. $^{\dagger}$ indicates the streaming BigCodec reproduced by the TS3-Codec authors.}
    \begin{tabular}{lccccccc}
      \toprule
      \textbf{Model} & \textbf{Bitrate} & \textbf{nParams} & \makecell{\textbf{Frame}\\\textbf{Rate}} & \makecell{\textbf{Token}\\\textbf{Rate}} & \makecell{\textbf{Codebook}\\\textbf{Layer}} & \textbf{Streaming} \\
      \midrule
      Ground Truth & - & - & - & - & - & - \\
      \midrule
      DAC & 1000 & 74.65M & 50 & 100 & 2 & $\times$ \\
      WavTokenizer & 900 & 80.9M & 75 & 75 & 1 & $\times$ \\
      SpeechTokenizer & 1000 & 103.7M & 50 & 100 & 2 & $\times$ \\
      SemantiCodec & 700 & 699.4M & 25 & 50 & 2 & $\times$ \\
      \midrule
      Encodec & 1500 & 14.85M & 75 & 150 & 2 & $\checkmark$ \\
      Mimi & 550 & 79.3M & 12 & 50 & 4 & $\checkmark$ \\
      Vocos & 1500 & 7.9M & 75 & 150 & 2 & $\checkmark$ \\
      SNAC & 984 & 19.8M & 46.88 & 82 & 3 & $\checkmark$ \\
      BigCodec$^{\dagger}$ & 1040 & 159.9M & 80 & 80 & 1 & $\checkmark$ \\
      TS3Codec & 850 & 203.6M & 50 & 50 & 1 & $\checkmark$ \\
      MagiCodec (Ours) & 850 & 209.7M & 50 & 50 & 1 & $\checkmark$ \\
      \bottomrule
    \end{tabular}
    \label{tab:computation}
\end{table}

\subsubsection{Tasks}
In addition to the reconstruction task, we conducted extensive downstream experiments to evaluate MagiCodec’s modelability, focusing on two main categories of tasks: generation and understanding.
We validate generative capability via zero-shot TTS, while the comprehension tasks encompass phone-level speech recognition, emotion recognition, and non-verbal detection.

\paragraph{Zero-shot TTS}
For TTS applications, using the codec’s outputs as intermediate speech representations allows us to assess whether the quantized features can drive a decoder to generate natural, coherent speech. Zero-shot TTS demands precise reproduction of speaker prosody, thereby revealing whether a codec preserves the information necessary for modeling rhythm and intonation. We evaluate multiple codecs on a zero-shot TTS task to determine whether MagiCodec can enhance downstream TTS systems or the performance of audio-based LLMs.

For the zero-shot TTS task, we extract discrete tokens from the LibriSpeech\maybeCite{panayotov2015librispeech} training set and evaluate the TTS model’s synthesis on utterances of 4 to 10 seconds from the LibriSpeech test-clean set.
Since multi-layer VQ codecs typically introduce additional modeling complexity and computational overhead in downstream language-modeling tasks, we restrict our baselines to single-layer codecs. We therefore include WavTokenizer and BigCodec as baselines, using their official pretrained weights.
We employ traditional TTS evaluation metrics to assess TTS performance, namely the aforementioned WER, PER, UTMOS, and SPK-SIM.

\paragraph{Phone-level Speech Recognition}
To more precisely assess the codec’s ability to preserve fine-grained speech details such as consonant and vowel transitions, we define the automatic speech recognition (ASR) task’s output units as phonemes rather than the more common word-level tokens (or Byte-Pair Encoding units). Phoneme-level recognition not only mitigates performance degradation due to out-of-vocabulary items but also reflects differences in intelligibility more sensitively after quantization through the phoneme error rate (PER).

For the phoneme-level speech recognition evaluation, discrete tokens are extracted from the LibriSpeech training set and performance is measured on utterances of 4 to 15 seconds drawn from the LibriSpeech test-clean set.
Only single-layer codecs are used as baselines, specifically WavTokenizer and BigCodec with their official pretrained weights.
PER is adopted as the evaluation metric to quantify recognition accuracy.

\paragraph{Emotion and Nonverbal Detection}
In addition to lexical sequences and phoneme-level information, we also aim to evaluate the capacity of codec tokens to capture a variety of acoustic cues beyond semantics. To this end, we conduct classification tasks for both emotion recognition and nonverbal detection. These paralinguistic cues represent performance factors that cutting-edge large language models are increasingly prioritizing~\maybeCite{}.

For the emotion classification task, we adopt the official training and testing splits of the English subset of the ESD dataset \maybeCite{zhou2022emotional}. The training set comprises 3,000 utterances from ten native speakers, while the test set contains 300 utterances drawn from speakers disjoint from those in the training set. The ESD corpus provides balanced coverage of five emotion categories (neutral, happiness, anger, sadness, and surprise), and each audio sample has an average duration of 2.7 seconds.
In the non-verbal detection evaluation, we use the VocalSound 16 kHz dataset \maybeCite{gong2022vocalsound}, which contains approximately 20k audio samples spanning six categories of non speech vocalizations, including laughter, sighs, coughs, throat clearing, sneezes and sniffs. Following the official data split, the training subset comprises 15,570 recordings while the test subset comprises 3,594 recordings. 
Alongside BigCodec and WavTokenizer we also employed DAC as a baseline, using each model’s official pretrained weights to extract discrete tokens and training downstream models with a single-layer VQ configuration.

\paragraph{Emotion and Nonverbal Detection}
In addition to lexical sequences and phoneme-level information, we also aim to evaluate the capacity of codec tokens to capture a variety of acoustic cues beyond semantics. To this end, we conduct classification tasks for both emotion recognition and nonverbal detection. These paralinguistic cues represent performance factors that cutting-edge large language models are increasingly prioritizing\maybeCite{}.

For the emotion classification task, we adopt the official training and testing splits of the English subset of the ESD dataset\maybeCite{zhou2022emotional}. The training set comprises 3,000 utterances from ten native speakers, while the test set contains 300 utterances drawn from speakers disjoint from those in the training set. The ESD corpus provides balanced coverage of five emotion categories (neutral, happiness, anger, sadness, and surprise), and each audio sample has an average duration of 2.7 seconds.

In the nonverbal detection evaluation, we use the VocalSound 16~kHz dataset\maybeCite{gong2022vocalsound}, which contains approximately 20,000 audio samples spanning six categories of non-speech vocalizations, including laughter, sighs, coughs, throat clearing, sneezes, and sniffs. Following the official data split, the training subset comprises 15,570 recordings, while the test subset comprises 3,594 recordings.

Alongside BigCodec and WavTokenizer, we also employ DAC as a baseline, using each model’s official pretrained weights to extract discrete tokens and training downstream models with a single-layer VQ configuration.

\subsubsection{Implementation Details}
\label{sec:imple}

All models were trained using 16 NVIDIA A100 80GB GPUs. Audio was uniformly resampled to 16~kHz, and each training sample was a randomly cropped, fixed-length 10-second segment to enhance dataset diversity. The batch size was adjusted according to model size and GPU memory constraints to achieve optimal hardware utilization.

Optimization was performed using the AdamW\maybeCite{loshchilov2017decoupled} algorithm with $\beta_1 = 0.8$, $\beta_2 = 0.99$, and $\epsilon = 1 \times 10^{-9}$ for robust convergence. The learning rates for the generator and discriminator were initially set to $1 \times 10^{-4}$, annealed to $1 \times 10^{-5}$ via a cosine schedule with a 1,000-step warmup phase. Training was conducted for a total of 100,000 steps.

For zero-shot TTS and ASR tasks, we trained the models from scratch using the official open-source GPT-2\footnote{\url{https://huggingface.co/openai-community/gpt2}}\maybeCite{radford2019language}. The model was trained on 8 NVIDIA A100 80GB GPUs.
We used a 12-layer, 12-head GPT-2 backbone with a 768-dimensional hidden size.
All dropout probabilities were fixed at 0.1.
Optimization was performed using AdamW ($\beta_1=0.9$, $\beta_2=0.999$).
The learning rate was set to $5\times10^{-4}$. After 5,000 warm-up steps, the rate followed a cosine annealing schedule that decayed smoothly to zero over the remaining updates. We trained the model for 20 epochs, with a batch size of 32.

For emotion and nonverbal detection tasks, we trained a BERT\maybeCite{devlin2019bert} model with the same training hyperparameters as above.

\subsection{Experimental Results}

\begin{table}[t]
  \centering
  \caption{Comparison of reconstruction ability between different codec models around 1000 bps and 50 token per second. $^{\dagger}$ indicates results taken from the TS3-Codec paper. Note that models marked with $^{\dagger}$ are trained on the same corpus as MagiCodec.}
  \resizebox{\textwidth}{!}{%
    \begin{tabular}{lcccccccc}
      \toprule
      \textbf{Model} & \textbf{WER}↓ & \textbf{PER}↓ & \textbf{STOI}↑ & \textbf{PESQ}↑ & \textbf{ViSQOL}↑ & \textbf{UTMOS}↑ & \textbf{SPK-SIM}↑ & \textbf{Streaming} \\
      \midrule
      Ground Truth & 1.85 & 0.79 & 1.00 & 4.64 & 5.00 & 4.09 & 1.00 & - \\
      \midrule
      DAC & 10.68 & 6.47 & 0.73 & 1.13 & 2.85 & 1.29 & 0.32 & $\times$ \\
      WavTokenizer & 3.75 & 1.95 & 0.90 & 2.13 & 3.95 & 3.79 & 0.66 & $\times$ \\
      SpeechTokenizer & 3.61 & 1.84 & 0.77 & 1.21 & 3.06 & 2.32 & 0.33 & $\times$ \\
      SemantiCodec & 4.61 & 2.55 & 0.86 & 1.79 & 3.83 & 2.93 & 0.61 & $\times$ \\
      \midrule
      Encodec & 4.22 & 2.15 & 0.85 & 1.56 & 3.59 & 1.58 & 0.60 & $\checkmark$ \\
      Mimi & 4.58 & 2.46 & 0.85 & 1.65 & 3.48 & 3.07 & 0.50 & $\checkmark$ \\
      Vocos & 4.32 & 2.39 & 0.89 & 1.96 & 3.79 & 3.04 & 0.63 & $\checkmark$ \\
      SNAC & 3.43 & 1.73 & 0.89 & 2.09 & 3.85 & 3.49 & 0.66 & $\checkmark$ \\
      BigCodec$^{\dagger}$ & 3.80 & - & 0.91 & 2.17 & - & 3.73 & 0.65 & $\checkmark$ \\
      TS3Codec$^{\dagger}$ & 3.60 & - & 0.91 & 2.23 & - & 3.84 & 0.68 & $\checkmark$ \\
      MagiCodec (Ours) & \textbf{3.16} & \textbf{1.63} & \textbf{0.93} & \textbf{2.56} & \textbf{4.15} & \textbf{4.18} & \textbf{0.76} & $\checkmark$ \\
      \bottomrule
    \end{tabular}
  }
\label{tab:recon}
\end{table}

\subsubsection{Reconstruction Quality}
Table~\ref{tab:recon} presents a comprehensive comparison of MagiCodec against several state-of-the-art neural audio codecs at a similar bitrate (\textasciitilde850–1000 bps) and token rate (50 tokens per second). All metrics reported are related to reconstruction quality. From the results, MagiCodec demonstrates clear advantages in multiple aspects:

1) \textbf{Speech Content Fidelity.} 
MagiCodec achieves the lowest Word Error Rate (WER) of 3.155 and Phoneme Error Rate (PER) of 1.634 among all neural codecs at comparable bitrate, significantly outperforming strong baselines such as BigCodec (WER 3.800) and TS3Codec (WER 3.600). This indicates that MagiCodec’s discrete representations preserve the linguistic content of speech more accurately, attributed to its noise-injected codebook optimization (Sec.3.2).

2) \textbf{Perceptual Quality and Intelligibility.}
In terms of perceptual metrics, MagiCodec attains a PESQ score of 2.562 and STOI of 0.925, surpassing all listed neural codecs by a noticeable margin. These improvements reflect higher perceived speech quality and intelligibility, approaching the natural speech baseline (PESQ 4.640, STOI 1.000). The VISQOL score of 4.147 further confirms MagiCodec’s capability in preserving fine-grained acoustic details, contributing to a more natural listening experience.

3) \textbf{Speaker Similarity and Naturalness.}
MagiCodec achieves the highest speaker similarity score (SPK-SIM = 0.762) and a leading naturalness measure (UTMOS = 4.183) among all compared models. This demonstrates that our codec effectively maintains speaker identity and prosodic characteristics during reconstruction.

4) \textbf{Streaming Capability.}
Despite its strong performance, MagiCodec maintains a moderate model size (209.7M parameters) and supports streaming inference with a single-layer codebook architecture. Compared to larger or multi-layer models such as SemanticCodec (699.4M parameters) and BigCodec (159.9M parameters), MagiCodec achieves a better trade-off between reconstruction quality and computational efficiency. Its streaming capability further enables real-time deployment scenarios.

\subsubsection{Generative modelability}

\begin{table}[t]
  \centering
  \caption{Zero-Shot TTS evaluation of various single-layer codecs to demonstrate their generative modelability on the LibriSpeech test-clean set.}
    \begin{tabular}{l c c c c c c}
    \toprule
    Model Name & Bitrate & WER$\downarrow$ & PER$\downarrow$ & UTMOS$\uparrow$  & SPK SIM$\uparrow$ & Streaming\\
    \midrule
    Ground Truth & - & 1.44 & 0.62 & 4.07 & 1.00 & - \\
    \midrule
    BigCodec & 1040 & 6.49 & 4.07 & 4.18 & \textbf{0.67} & $\times$ \\
    WavTokenizer & 900 & 3.83 & 1.91 & 3.95 & 0.54 & $\checkmark$ \\
    MagiCodec (Ours) & 850 & \textbf{3.30} & \textbf{1.71} & \textbf{4.27} & 0.61 & $\checkmark$ \\
    \bottomrule
    \end{tabular}
  \label{tab:tts}
\end{table}

Table \ref{tab:tts} shows the results of zero-shot TTS task. MagiCodec achieves the lowest word error rate (WER = 3.30 \%) and phoneme error rate (PER = 1.71 \%) at just 850 bps, while also registering the highest naturalness score (UTMOS = 4.27). This represents a substantial improvement over the single-layer quantizer WavTokenizer and the higher-bitrate, non-streaming BigCodec. Although BigCodec edges out MagiCodec slightly in speaker similarity, that advantage comes with significantly greater bitrate overhead and non-streaming latency. MagiCodec tokens make the TTS model more predictable, thereby enabling it to lead in both content accuracy and naturalness.

\begin{table}[t]
  \centering
  \caption{Results of phone-level speech recognition for various codecs.}
    \begin{tabular}{l c c c c c c}
    \toprule
    Model Name & PER$\downarrow$ \\
    \midrule
    BigCodec & 8.0 \\
    WavTokenizer & 13.1 \\
    MagiCodec (Ours) & 7.7 \\
    \bottomrule
    \end{tabular}
  \label{tab:asr}
\end{table}
\begin{table}[t]
  \centering
  \caption{Performance comparison of various codecs on downstream tasks of sentiment classification and non-verbal detection. Each model was trained 10 times, and the standard deviation of each evaluation metric is shown as a superscript to the mean value.}
  \begin{tabular}{l c c  c c}
    \toprule
    & \multicolumn{2}{c}{Sentiment Classification} 
    & \multicolumn{2}{c}{Non-verbal Detection} \\
    \cmidrule(lr){2-3} \cmidrule(lr){4-5}
        Model Name & ACC$\uparrow$ & F1$\uparrow$ & ACC$\uparrow$ & F1$\uparrow$ \\
    \midrule
        DAC             & $0.54_{0.006}$ & $0.54_{0.006}$ & $0.59_{0.006}$ & $0.59_{0.006}$ \\
        BigCodec        & $0.59_{0.010}$ & $0.59_{0.009}$ & $0.51_{0.007}$ & $0.51_{0.007}$ \\
        WavTokenizer & $0.62_{0.009}$ & $0.62_{0.008}$ & $0.59_{0.006}$ & $0.59_{0.006}$ \\
    \midrule
        MagiCodec (Ours)       & $\mathbf{0.70}_{0.017}$ & $\mathbf{0.70}_{0.016}$ & $\mathbf{0.63}_{0.007}$ & $\mathbf{0.63}_{0.007}$ \\
    \midrule
    \bottomrule
  \end{tabular}
  \label{tab:comprehension}
\end{table}

\subsubsection{Comprehension capability}

We assess the phone-level modeling capacity of each codec by training an ASR model to predict phoneme sequences from codec tokens. As shown in Table \ref{tab:asr}, MagiCodec achieves the lowest phone error rate (PER) of 7.7\%, outperforming BigCodec (8.0\%) and substantially improving over WavTokenizer (13.1\%). This reduction in PER indicates that MagiCodec’s discrete representations preserve finer-grained phonetic information.

Next, we evaluate the comprehension capability on sentiment classification and non-verbal detection. we report mean accuracy and F1 (with standard deviations over 10 runs) in Table \ref{tab:comprehension}. MagiCodec again leads, achieving 70\% accuracy and F1 on sentiment classification and 63\% accuracy and F1 on non-verbal detection. By comparison, WavTokenizer attains 62\% on both metrics for sentiment and 59\% for non-verbal detection, while BigCodec lags further behind.

Taken together, these results demonstrate that MagiCodec’s single-layer quantization not only excels at retaining phonetic detail (lower PER) but also encodes richer semantic and paralinguistic cues, thereby enhancing modelability on diverse downstream tasks.

\begin{table}[t]
  \centering
  \caption{Ablation study on reconstruction metrics with MagiCodec under different mask ratios, token rates, and encoder sizes.}
  \resizebox{\textwidth}{!}{
  \begin{tabular}{lccccccc}
    \toprule
    & \multicolumn{7}{c}{\textbf{Reconstruction}} \\
    \cmidrule(lr){2-8}
    \textbf{Model} & WER$\downarrow$ & PER$\downarrow$ & STOI$\uparrow$ & PESQ$\uparrow$ & ViSQOL$\uparrow$ & UTMOS$\uparrow$ & SPK SIM$\uparrow$ \\
    \midrule
    50Hz mask $0\%$ & 3.34 & 1.77 & 0.93 & 2.56 & 4.16 & 4.17 & 0.77 \\
    50Hz mask $10\%$ & 3.22 & 1.68 & 0.93 & 2.57 & 4.17 & 4.18 & 0.77 \\
    50Hz mask $20\%$ & 3.16 & 1.63 & 0.93 & 2.56 & 4.15 & 4.18 & 0.76 \\
    50Hz mask $30\%$ & 3.17 & 1.62 & 0.93 & 2.56 & 4.16 & 4.17 & 0.76 \\
    \midrule
    25Hz & 6.59 & 3.83 & 0.88 & 1.90 & 3.85 & 3.59 & 0.61 \\
    100Hz & 2.23 & 1.04 & 0.95 & 3.00 & 4.34 & 4.19 & 0.87 \\
    \bottomrule
  \end{tabular}
  }
\label{tab:ablation:main}
\end{table}

\begin{table}[t]
  \centering
  \caption{Ablation study on TTS, emotion and non-verbal tasks across model variants.}
  \resizebox{\textwidth}{!}{
    \begin{tabular}{lccccccccc}
      \toprule
      & \multicolumn{4}{c}{\textbf{TTS}} 
      & \multicolumn{2}{c}{\textbf{Emotion}} 
      & \multicolumn{2}{c}{\textbf{Non-verbal}} \\
      \cmidrule(lr){2-5}\cmidrule(lr){6-7}\cmidrule(lr){8-9}
      \textbf{Model} & WER$\downarrow$ & PER$\downarrow$ & UTMOS$\uparrow$ & SPK SIM$\uparrow$
                    & ACC$\uparrow$ & F1$\uparrow$
                    & ACC$\uparrow$ & F1$\uparrow$ \\
      \midrule
      50Hz mask $0\%$          & 5.51 & 3.26 & 4.24 & 0.62 & $0.68_{0.02}$ & $0.68_{0.02}$ & $0.61_{0.01}$ & $0.62_{0.01}$ \\
      50Hz mask $10\%$         & 3.57 & 1.88 & 4.27 & 0.62 & $0.69_{0.02}$ & $0.69_{0.02}$ & $\mathbf{0.63}_{0.01}$ & $\mathbf{0.63}_{0.01}$ \\
      50Hz mask $20\%$         & 3.37 & \textbf{1.70} & 4.28 & 0.62 & $0.67_{0.03}$ & $0.67_{0.03}$ & $0.62_{0.01}$ & $\mathbf{0.63}_{0.01}$ \\
      50Hz mask $30\%$         & \textbf{3.30} & 1.71 & \textbf{4.28} & 0.62 & $\mathbf{0.70}_{0.02}$ & $\mathbf{0.70}_{0.02}$ & $\mathbf{0.63}_{0.01}$ & $\mathbf{0.63}_{0.01}$ \\
      \midrule
      25Hz                & 5.47 & 2.77 & 3.74 & 0.49 & $0.54_{0.02}$ & $0.54_{0.02}$ & $0.57_{0.01}$ & $0.57_{0.01}$ \\
      100Hz               & --   & --   & --   & --   & $0.58_{0.03}$ & $0.58_{0.03}$ & $0.57_{0.01}$ & $0.58_{0.01}$ \\
      \bottomrule
    \end{tabular}
  }
\label{tab:ablation:downstream}
\end{table}
\subsection{Ablation Study}
We conducted the ablation experiments using the same experimental settings as in the main experiments.
The reconstruction results are shown in Table~\ref{tab:ablation:main} and the downstream scores in Table~\ref{tab:ablation:downstream}. 

\paragraph{Mask ratio.}
Increasing the proportion of masked frames yields consistent gains across almost all metrics. On the core reconstruction benchmarks, WER drops from $3.34$ to $3.16$ as the mask ratio rises to $20\%$, and then plateaus ($3.17$ at $30\%$). Similar monotonic improvements appear for PER and the perceptual metrics (PESQ, ViSQOL, UTMOS), suggesting that moderate corruption encourages the encoder to form more robust, context-aware representations.

We can see that zero-shot TTS reaches its lowest WER at $30\%$ masking ($3.30$) and emotion recognition peaks at the same ratio with ${\rm ACC}=0.70$ and ${\rm F1}=0.70$. We hypothesize that hiding up to one third of the acoustic codes forces the quantizer to infer longer-range semantic structure—an effect reminiscent of the gestalt reasoning observed in MAE for images.

\paragraph{Token rate.}
Changing the token size trades temporal resolution against sequence length. Halving the rate to $25,\mathrm{Hz}$ degrades reconstruction severely (WER $6.59$) and harms every downstream task, confirming that information is simply discarded when tokens are too sparse. Conversely, doubling the rate to $100,\mathrm{Hz}$ pushes reconstruction to its best numbers (WER $2.23$, STOI $0.95$, PESQ $3.00$), but the longer sequences complicate autoregressive generation and thus does harm to downstream tasks. Emotion and non-verbal detection improve only marginally. Taken together, $50\mathrm{Hz}$ offers the best compromise between fidelity and modelability, aligning with prior work that caps useful temporal granularity near the frame rate of human speech perception.

We observe that moderate levels of masking consistently improve both the reconstruction performance and the generative usability of the codec, a trend that aligns with observations in masked image modeling. On the other hand, setting the token rate too low leads to the loss of essential phonetic information, while excessively high token rates offer little additional benefit for downstream tasks and may even introduce redundancy.

\begin{figure}[t]
    \centering
    \begin{subfigure}{0.32\linewidth}
        \includegraphics[width=\linewidth]{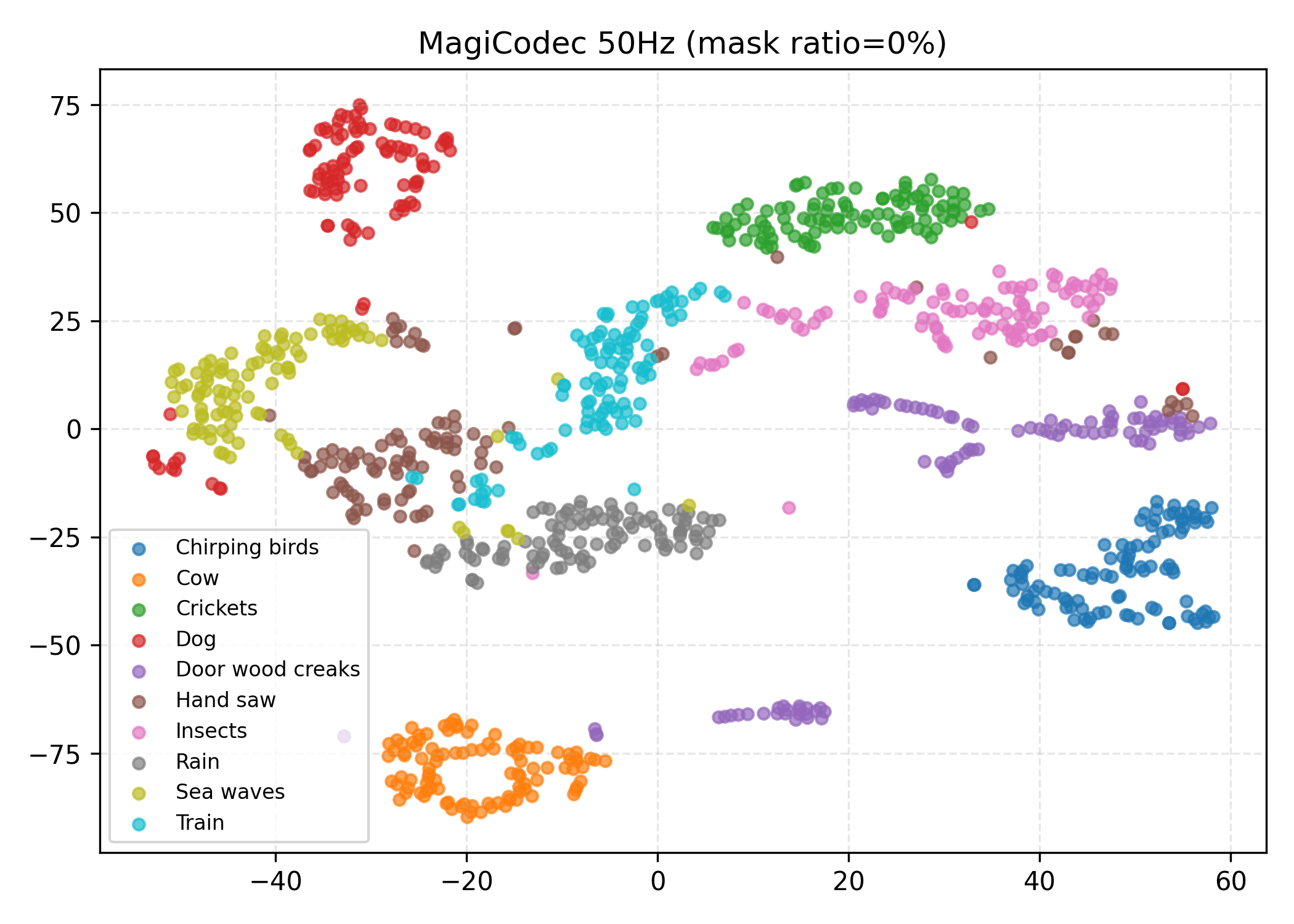}
        \caption{MagiCodec (mask 0\%)}
    \end{subfigure}
    \begin{subfigure}{0.32\linewidth}
        \includegraphics[width=\linewidth]{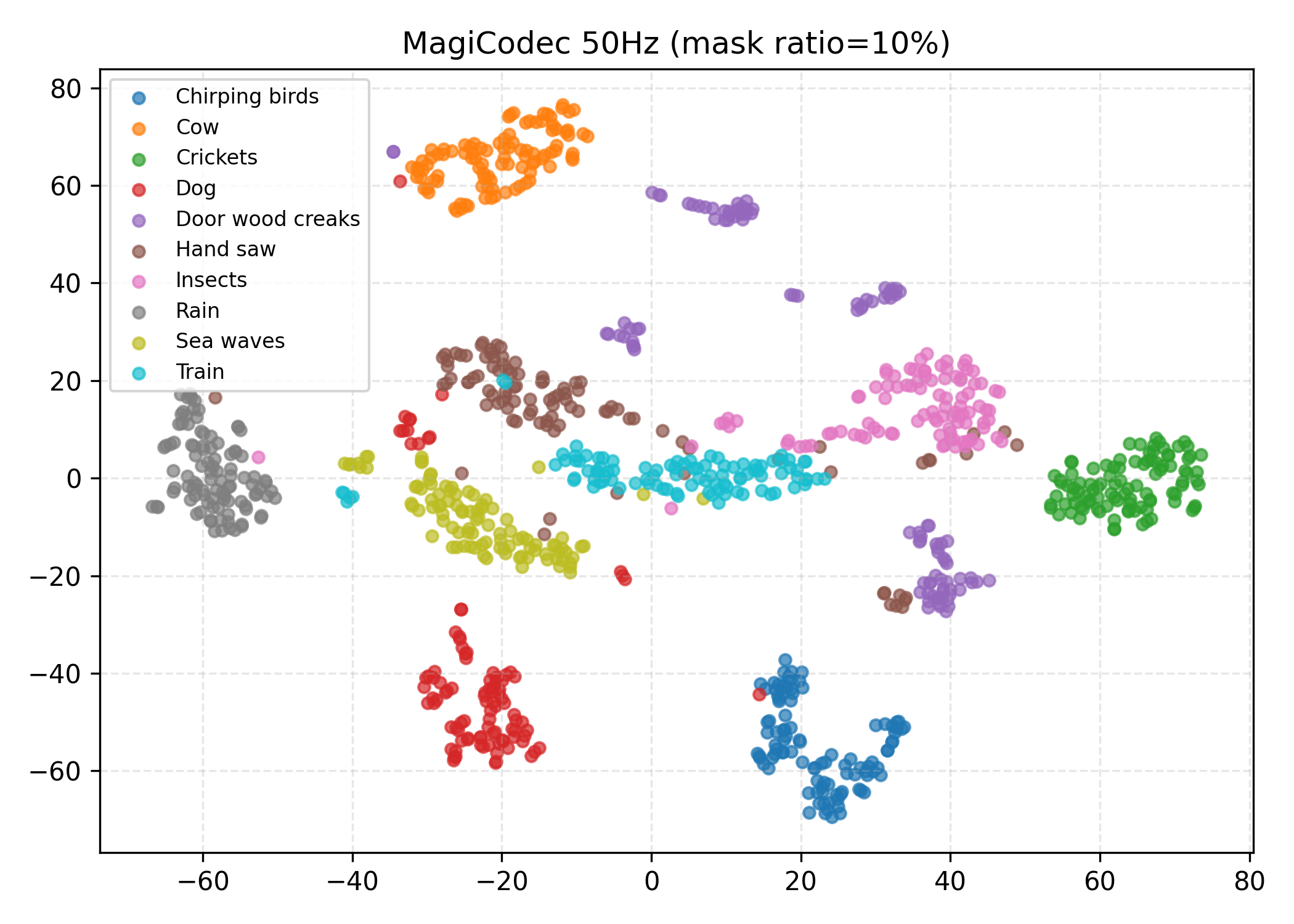}
        \caption{MagiCodec (mask 10\%)}
    \end{subfigure}
    \begin{subfigure}{0.32\linewidth}
        \includegraphics[width=\linewidth]{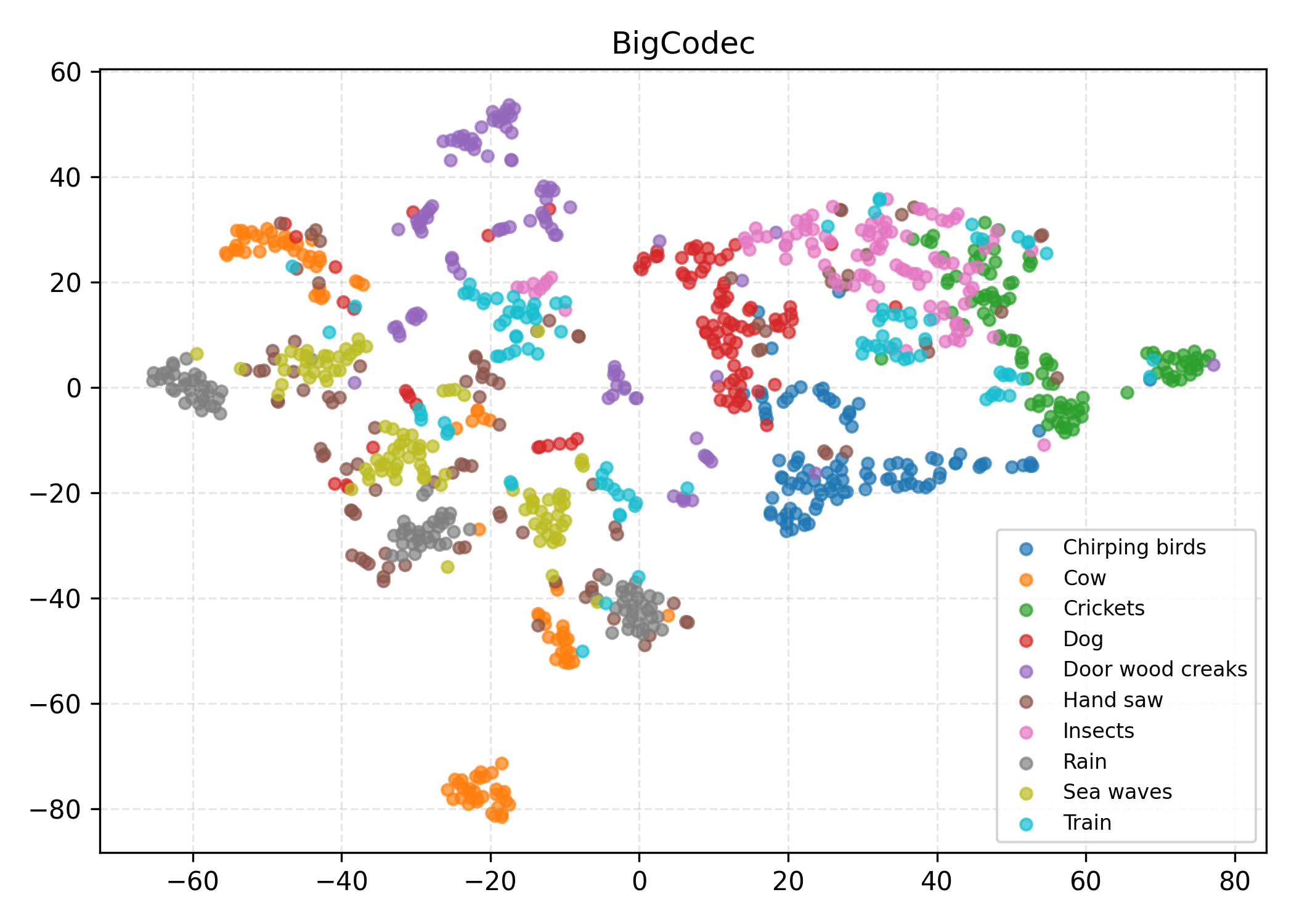}
        \caption{BigCodec}
    \end{subfigure}
    \vspace{3mm}
    \begin{subfigure}{0.32\linewidth}
        \includegraphics[width=\linewidth]{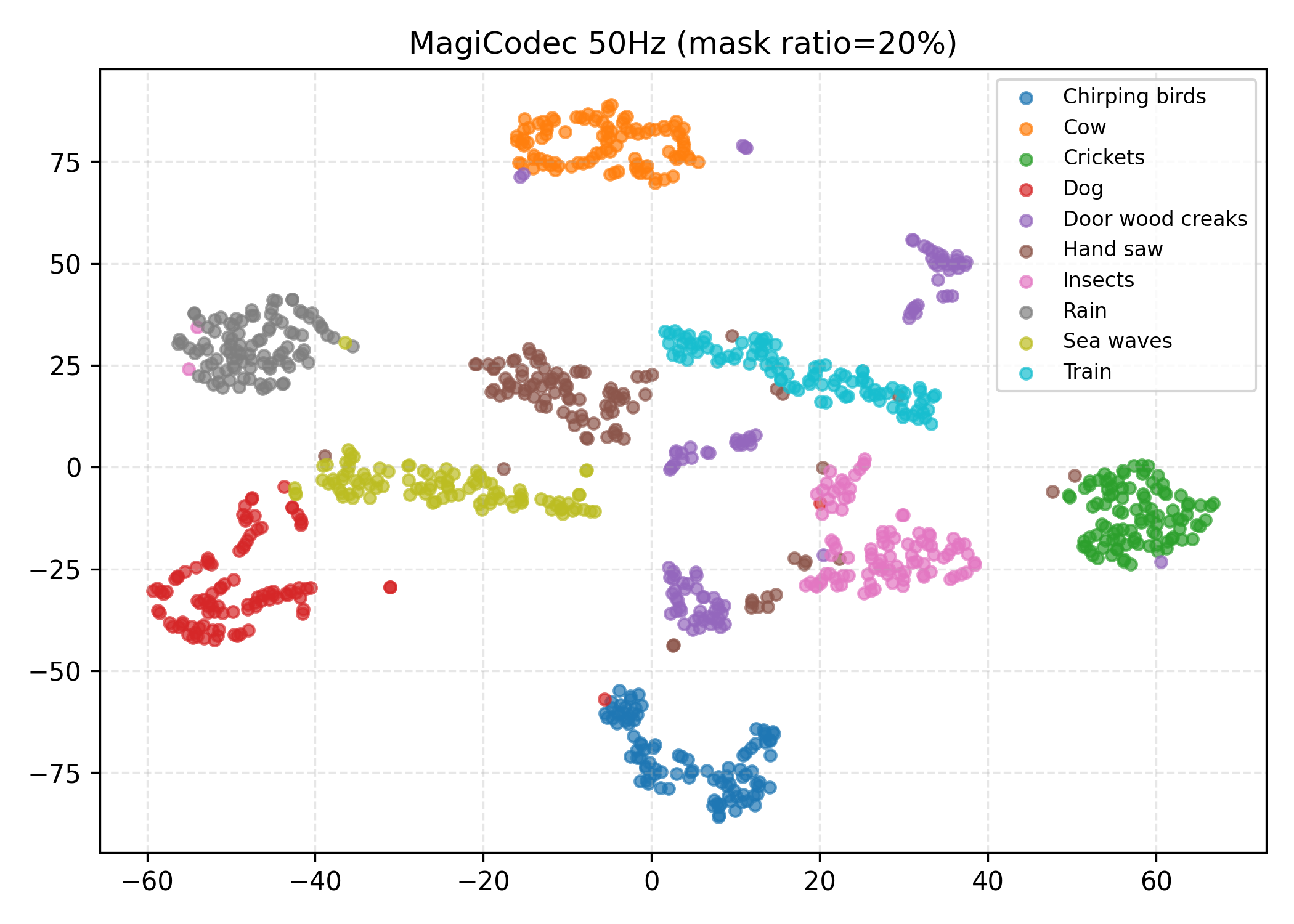}
        \caption{MagiCodec (mask 20\%)}
    \end{subfigure}
    \begin{subfigure}{0.32\linewidth}
        \includegraphics[width=\linewidth]{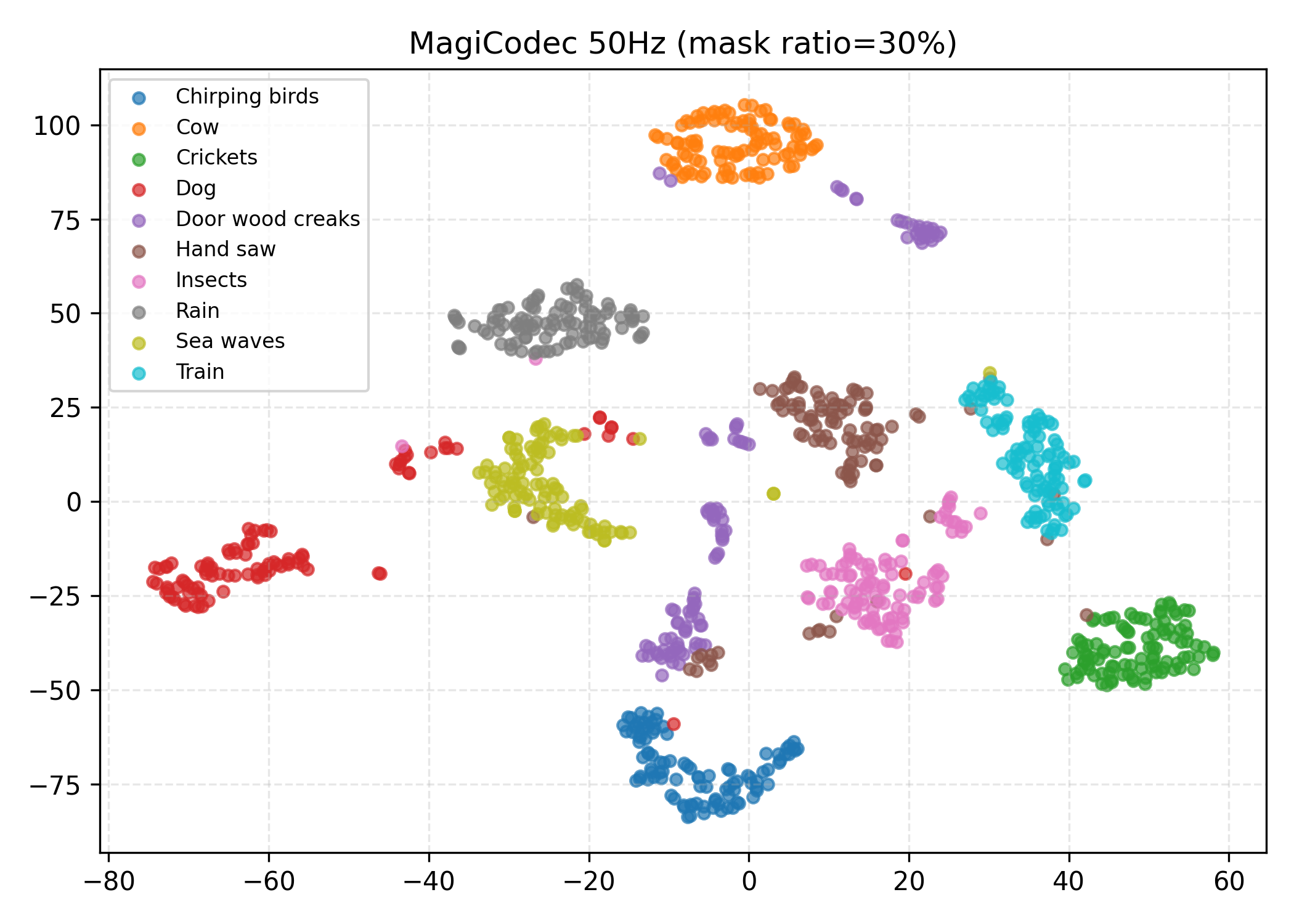}
        \caption{MagiCodec (mask 30\%)}
    \end{subfigure}
    \begin{subfigure}{0.32\linewidth}
        \includegraphics[width=\linewidth]{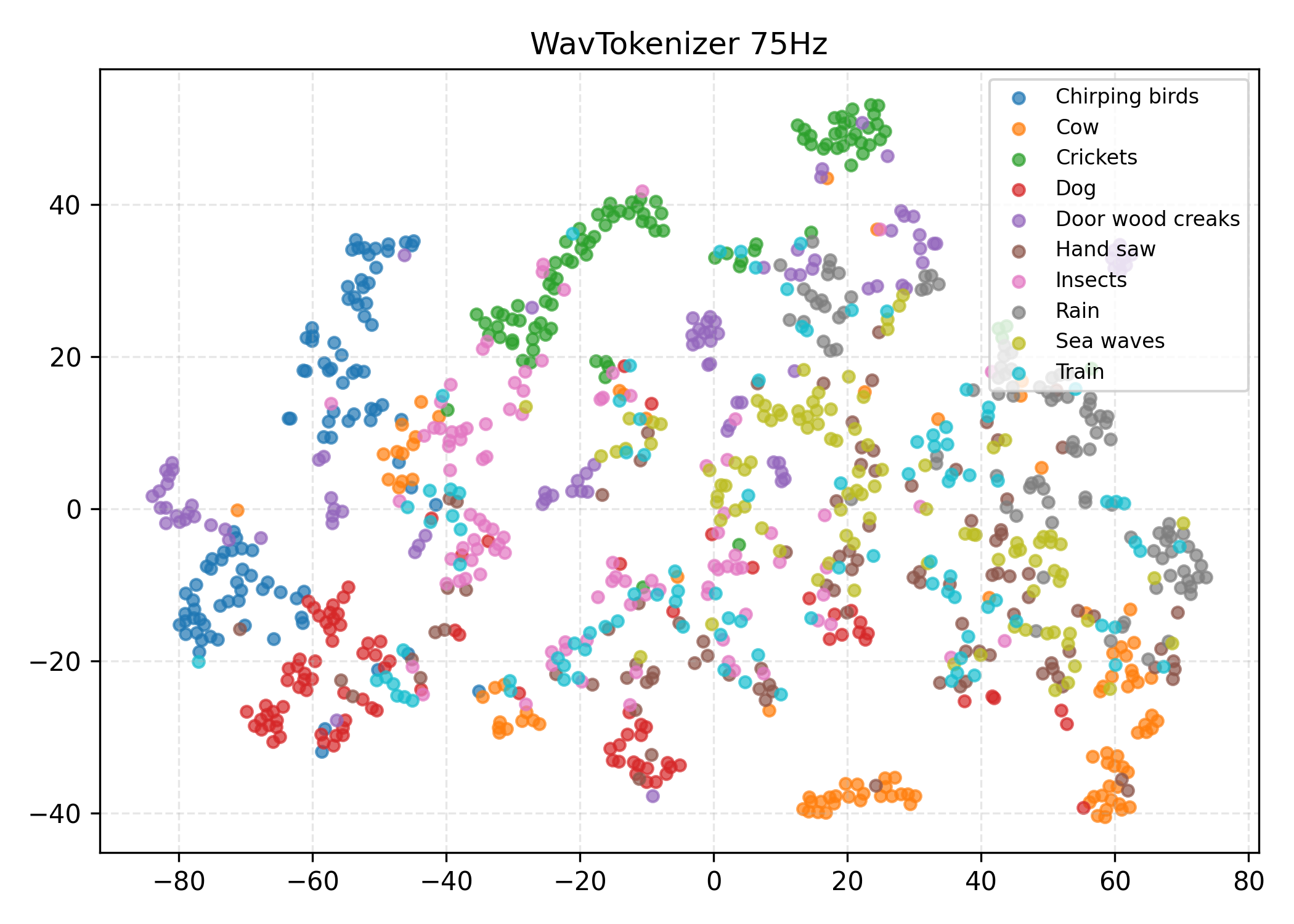}
        \caption{WavTokenizer}
    \end{subfigure}
    \caption{Visualization of the latent space using tSNE and 10 random classes in ESC-50 dataset.}
    \label{fig:tSNE}
\end{figure}
\paragraph{Latent Visualization}
To provide a more intuitive comparison of the encoding results from different models, we visualize the latent representations extracted by MagicCodec, BigCodec, and wavtokenizer. To this end, we employ t-SNE to project the high-dimensional latent spaces of these models onto a two-dimensional plane using the ESC-50 dataset \maybeCite{piczak2015dataset}. As shown in Figure \ref{fig:tSNE}, the latent representations produced by MagicCodec exhibit more distinct clustering in the two-dimensional space, with samples from the same audio class being grouped more closely together compared to the other models. In contrast, the latent spaces of BigCodec and wavtokenizer show less clear separation between different audio categories, with more overlap observed among classes.

Furthermore, we investigate the effect of varying the mask ratio in MagicCodec. Our experiments reveal that increasing the mask ratio leads to a more concentrated semantic distribution in the latent space, as evidenced by tighter and more compact clusters in the t-SNE visualization. This suggests that a higher mask ratio encourages the model to learn more abstract and semantically meaningful representations. Overall, these results demonstrate that MagicCodec not only achieves superior clustering performance in the latent space but also benefits from enhanced semantic structure as the mask ratio increases.

\paragraph{Token Distribution}

It is well-known that text tokens in natural language follow Zipf’s law~\maybeCite{stanisz2024complex,chan2024analyzing}, where a few high-frequency tokens dominate while many low-frequency tokens are sparse, reflecting rich semantic hierarchy. If audio tokens exhibit a similar Zipf distribution, it indicates strong semantic representation capability.

\begin{figure}[t]
    \centering
    \begin{subfigure}{0.32\linewidth}
        \includegraphics[width=\linewidth]{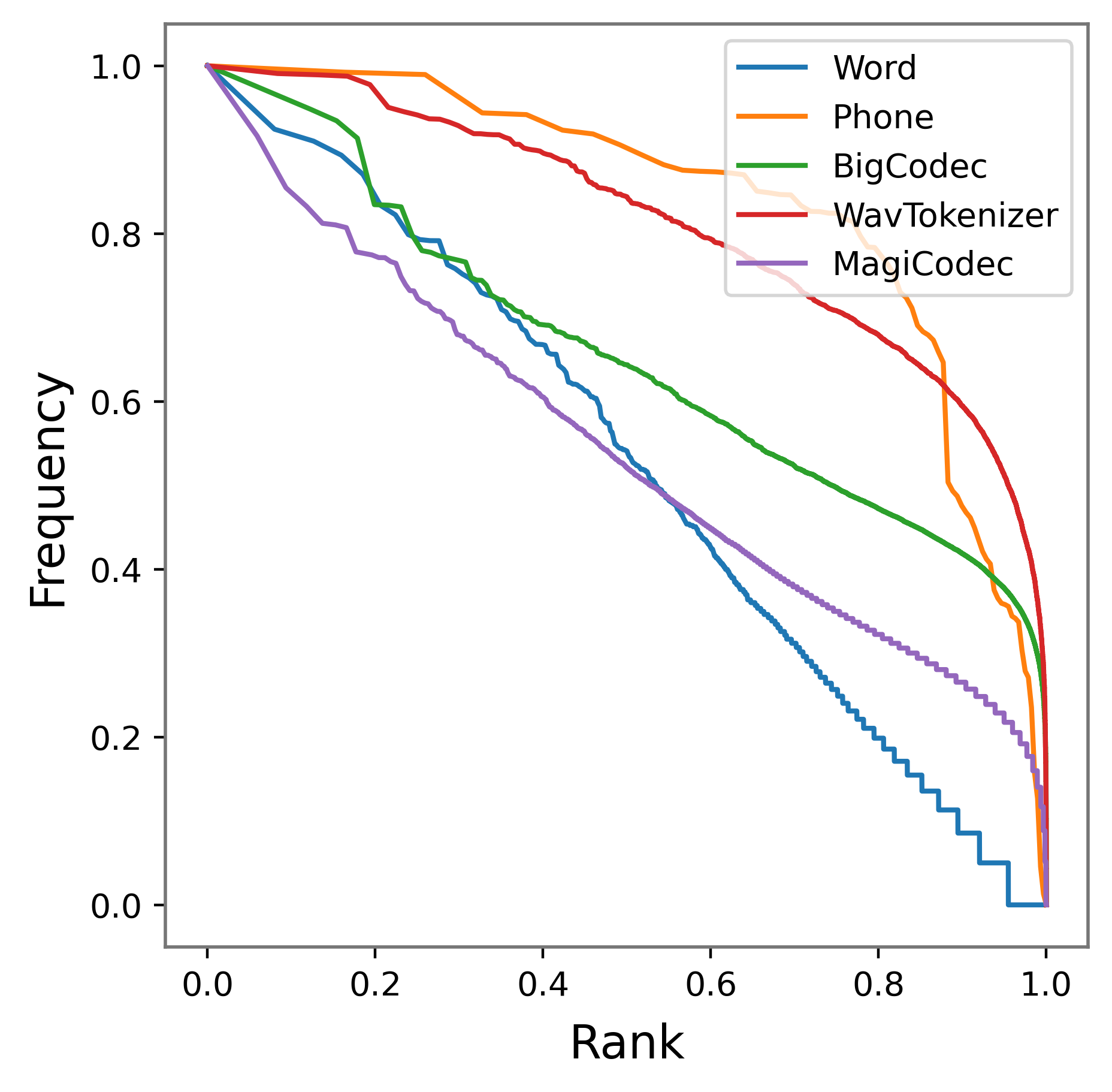}
        \caption{1-grams}
    \end{subfigure}
    \begin{subfigure}{0.32\linewidth}
        \includegraphics[width=\linewidth]{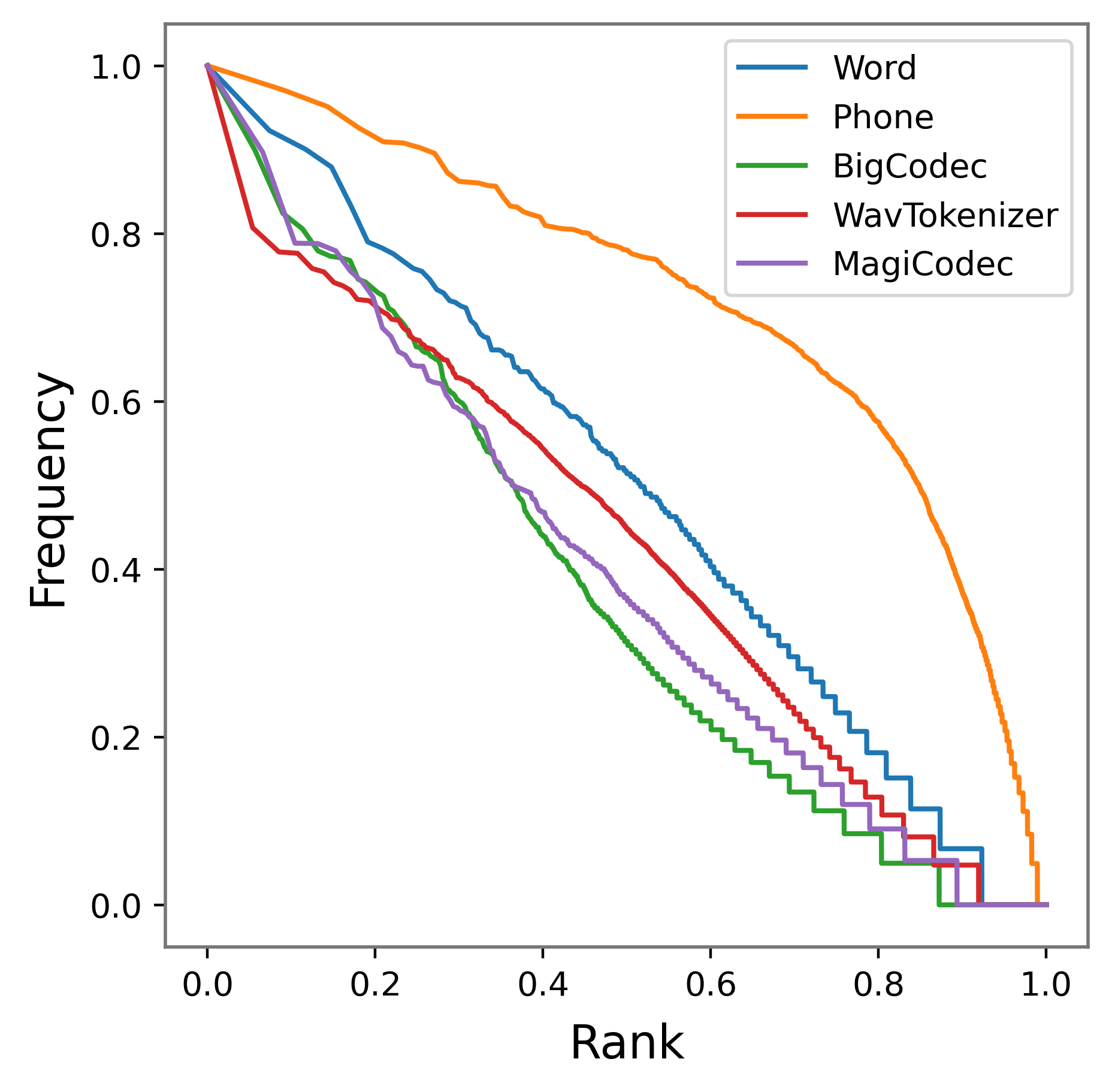}
        \caption{2-grams}
    \end{subfigure}
    \begin{subfigure}{0.32\linewidth}
        \includegraphics[width=\linewidth]{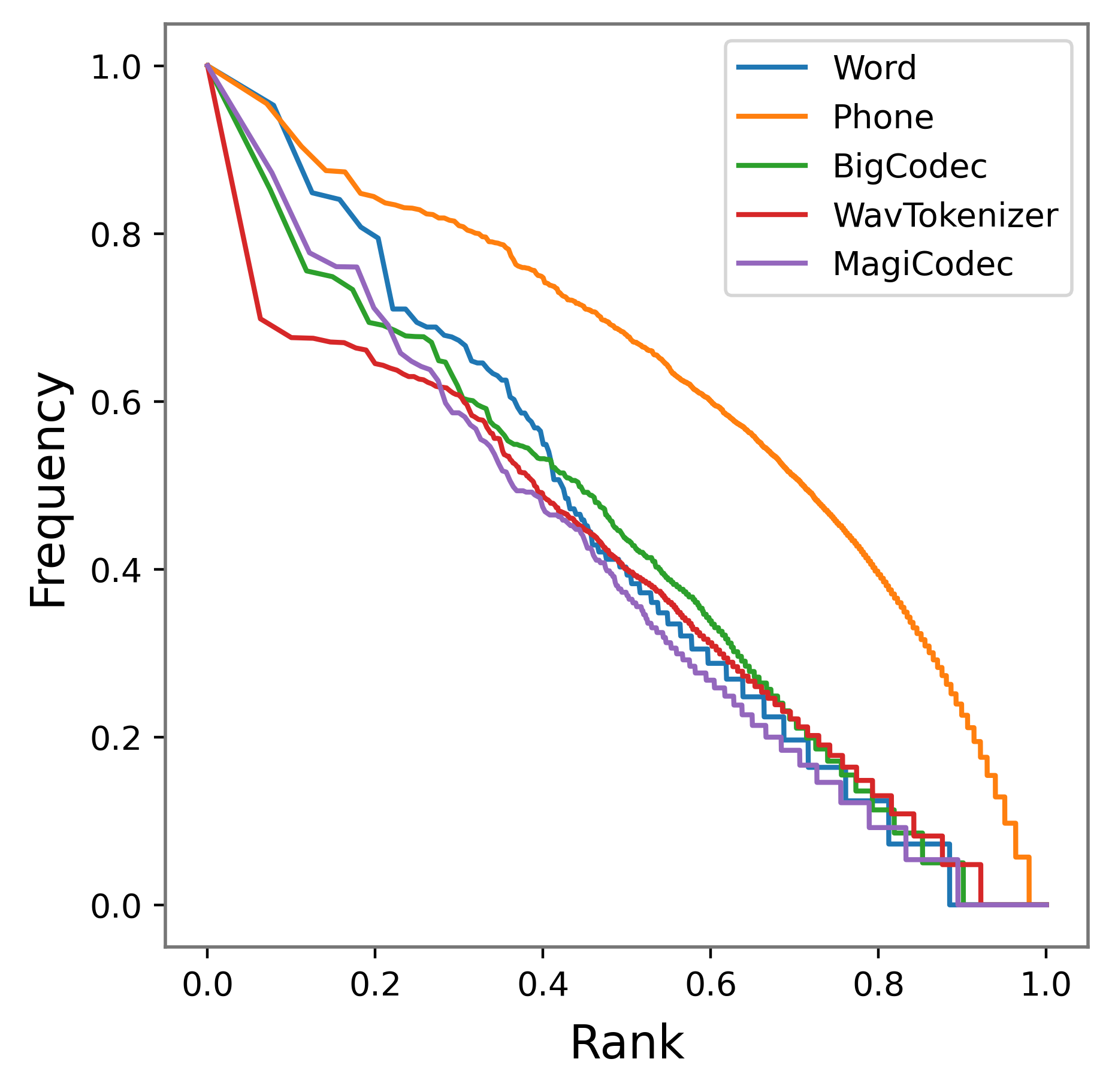}
        \caption{3-grams}
    \end{subfigure}
    \begin{subfigure}{0.32\linewidth}
        \includegraphics[width=\linewidth]{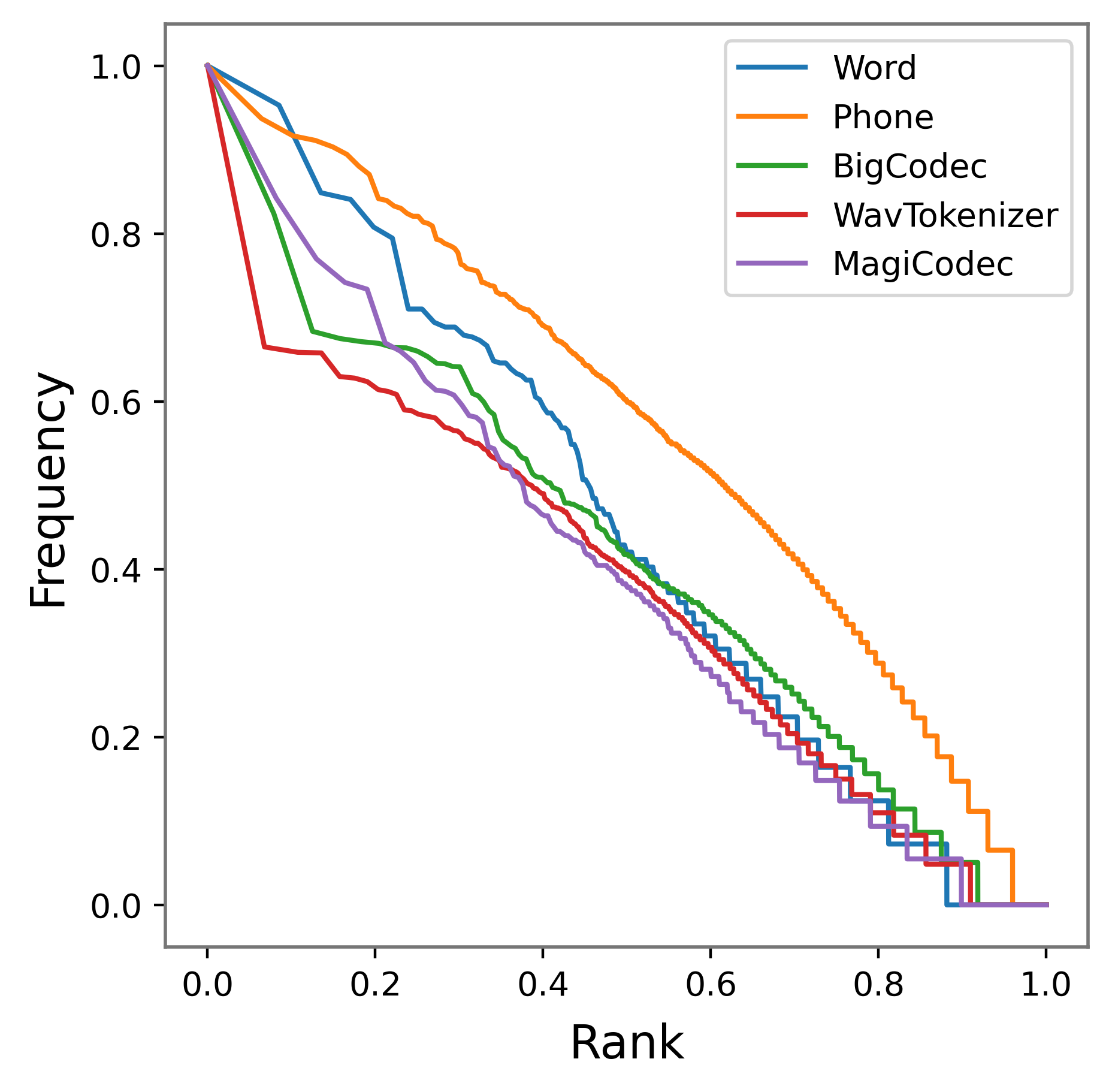}
        \caption{4-grams}
    \end{subfigure}
    \begin{subfigure}{0.32\linewidth}
        \includegraphics[width=\linewidth]{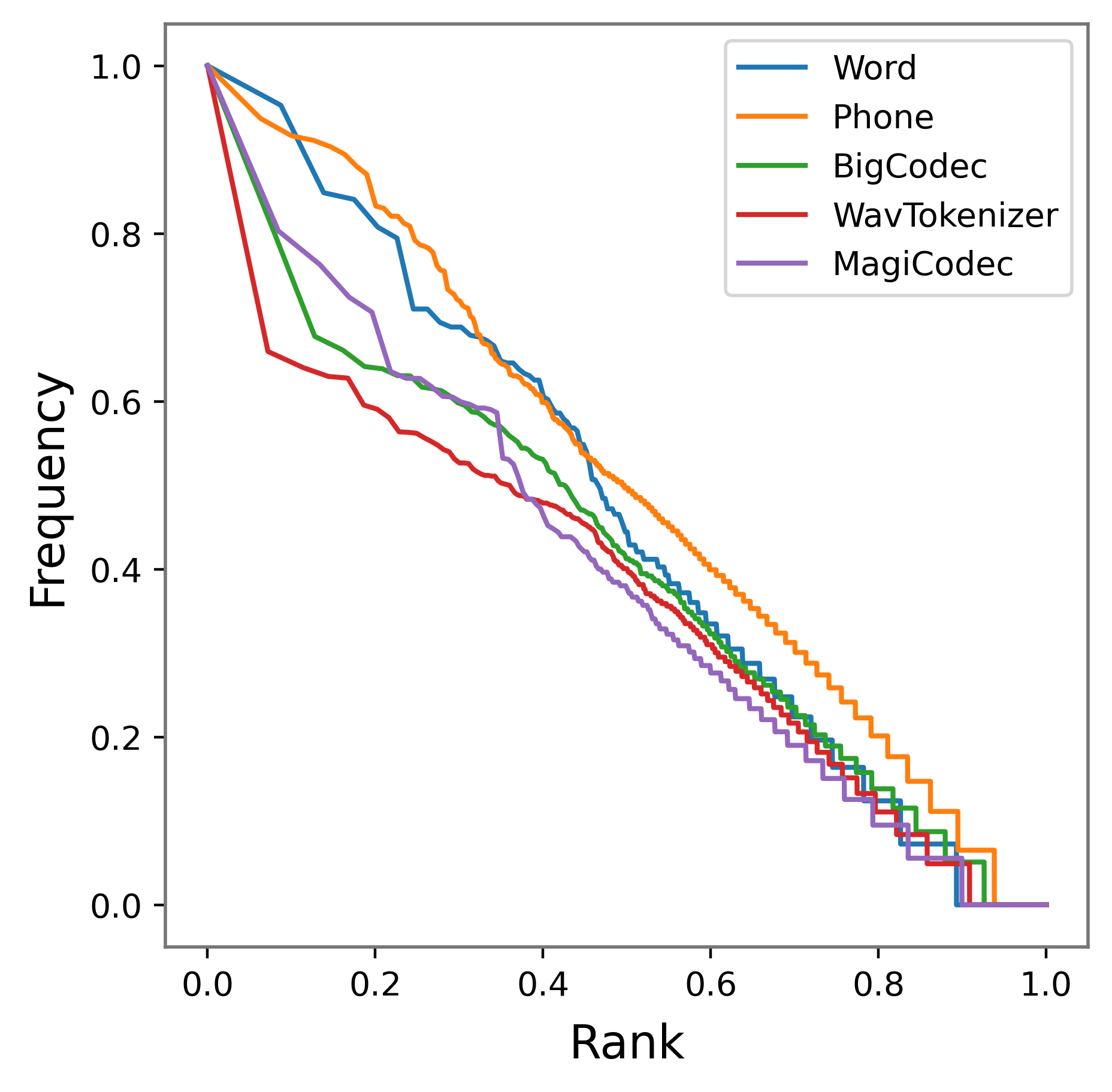}
        \caption{5-grams}
    \end{subfigure}
    \begin{subfigure}{0.32\linewidth}
        \includegraphics[width=\linewidth]{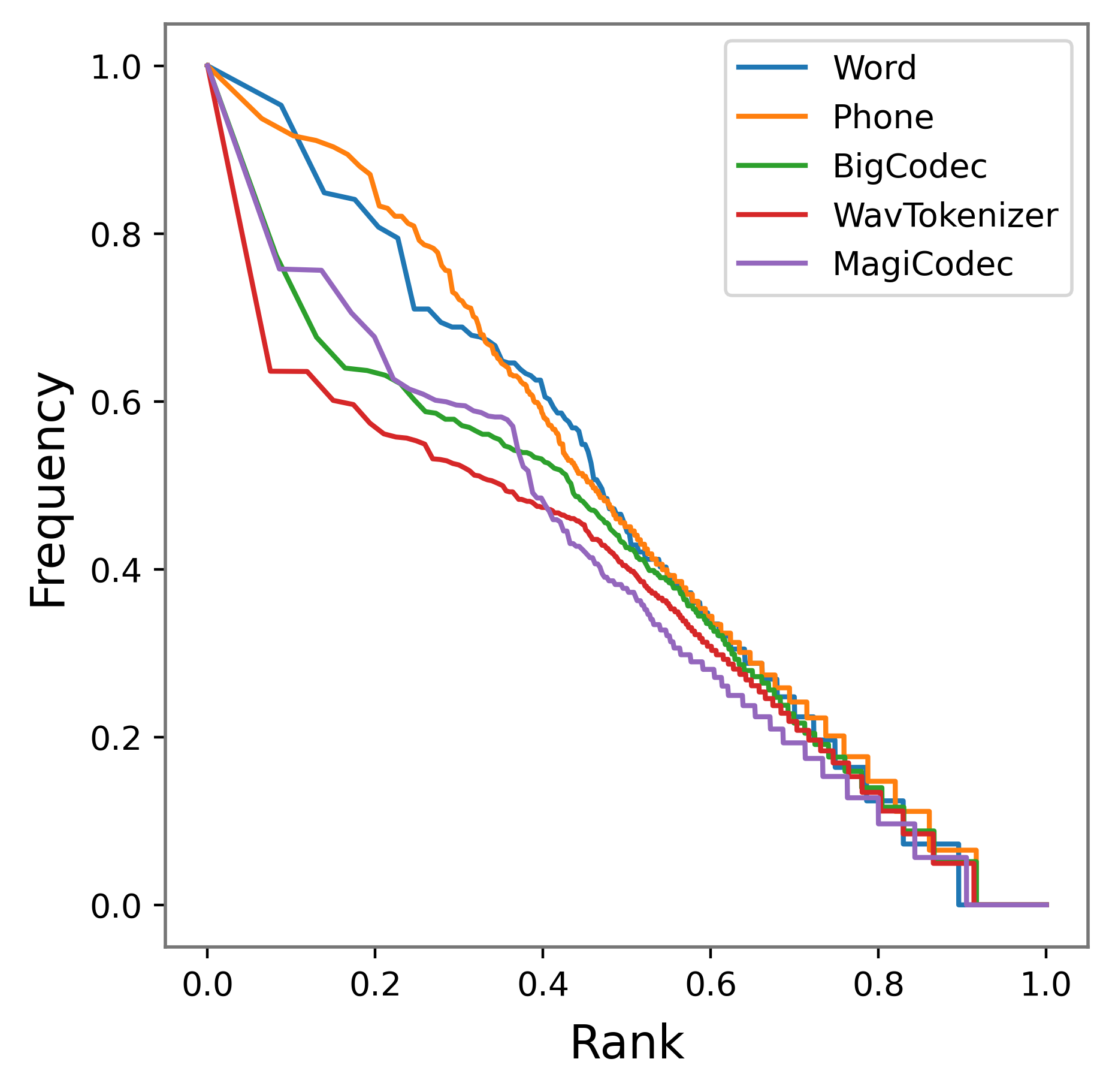}
        \caption{6-grams}
    \end{subfigure}
    \caption{Plots of normalized token log-frequency versus normalized log-rank for various codec models and natural languages across different n-gram levels. Hapax tokens (tokens appearing only once) have been excluded for all visualizations.}
    \label{fig:zipf}
\end{figure}

We conducted a visualization analysis in \ref{fig:zipf}, which shows normalized frequency versus rank for various token sets and $n$-grams ($n=1$ to $6$), including: 
1) text word tokens (semantic gold standard), 
2) phoneme-level tokens (less semantic content), 
3) existing audio tokenization methods, and 
4) the proposed MagiCodec.

And we have several observations: 
1) Word tokens display a clear power-law decay across all $n$-grams, consistent with natural language. 
2) Phone tokens have a flatter distribution, especially for 1- and 2-grams, indicating weaker semantic hierarchy. 
3) Existing audio tokens fall between phone and word tokens; as $n$ increases, their distributions approach that of words but remain less semantically rich. 
4) MagiCodec’s distribution closely matches word tokens across all $n$-grams, particularly for $n \geq 3$, suggesting strong semantic structure and contextual dependence in its representations.

\section{Conclusion}

This paper presents MagiCodec, a simple yet high-performance single-layer streaming audio codec. Compared to state-of-the-art models, MagiCodec improves both reconstruction quality and downstream modelability by generating discrete tokens that better capture semantic and paralinguistic information. A multistage training pipeline with noise injection and latent regularization enables MagiCodec to achieve superior results in both objective and downstream tasks, demonstrating its effectiveness and broad potential in neural audio processing.

\section{Discussion}
\paragraph{Limitation}
\label{sec:limitation}
Although MagiCodec achieves strong speech reconstruction and downstream task performance, the single-layer quantization may still limit the preservation of fine details in broadband audio such as music. Moreover, since training is conducted only on 16kHz English speech, the robustness of the codec in noisy conditions or at higher sampling rates remains untested.

\paragraph{Broader impacts}
\label{sec:broader}
The model is able to maintain high quality even at low bitrates, thereby reducing energy consumption during both training and inference. However, improved reconstruction capabilities may also facilitate unauthorized voice cloning or deepfakes. We encourage researchers to incorporate watermarking, detection tools, and clear usage policies when releasing downstream model weights and interfaces, and we urge the community to remain vigilant and monitor potential misuse.

\clearpage

\bibliography{references}

\begin{thebibliography}{51}
\providecommand{\natexlab}[1]{#1}
\providecommand{\url}[1]{\texttt{#1}}
\expandafter\ifx\csname urlstyle\endcsname\relax
  \providecommand{\doi}[1]{doi: #1}\else
  \providecommand{\doi}{doi: \begingroup \urlstyle{rm}\Url}\fi

\bibitem[Achiam et~al.(2023)Achiam, Adler, Agarwal, Ahmad, Akkaya, Aleman, Almeida, Altenschmidt, Altman, Anadkat, et~al.]{achiam2023gpt}
Josh Achiam, Steven Adler, Sandhini Agarwal, Lama Ahmad, Ilge Akkaya, Florencia~Leoni Aleman, Diogo Almeida, Janko Altenschmidt, Sam Altman, Shyamal Anadkat, et~al.
\newblock {GPT}-4 technical report.
\newblock \emph{arXiv preprint arXiv:2303.08774}, 2023.

\bibitem[Ai et~al.(2024)Ai, Jiang, Lu, Du, and Ling]{ai2024apcodec}
Yang Ai, Xiao-Hang Jiang, Ye-Xin Lu, Hui-Peng Du, and Zhen-Hua Ling.
\newblock {APCodec}: A neural audio codec with parallel amplitude and phase spectrum encoding and decoding.
\newblock \emph{IEEE/ACM Transactions on Audio, Speech, and Language Processing}, 2024.

\bibitem[Anastassiou et~al.(2024)Anastassiou, Chen, Chen, Chen, Chen, Chen, Cong, Deng, Ding, Gao, et~al.]{anastassiou2024seed}
Philip Anastassiou, Jiawei Chen, Jitong Chen, Yuanzhe Chen, Zhuo Chen, Ziyi Chen, Jian Cong, Lelai Deng, Chuang Ding, Lu~Gao, et~al.
\newblock {Seed-TTS}: A family of high-quality versatile speech generation models.
\newblock \emph{arXiv preprint arXiv:2406.02430}, 2024.

\bibitem[Bengio et~al.(2013)Bengio, L{\'e}onard, and Courville]{bengio2013estimating}
Yoshua Bengio, Nicholas L{\'e}onard, and Aaron Courville.
\newblock Estimating or propagating gradients through stochastic neurons for conditional computation.
\newblock \emph{arXiv preprint arXiv:1308.3432}, 2013.

\bibitem[Bishop(1995)]{bishop1995}
Chris~M. Bishop.
\newblock Training with noise is equivalent to tikhonov regularization.
\newblock \emph{Neural Computation}, 7\penalty0 (1):\penalty0 108--116, 1995.
\newblock \doi{10.1162/neco.1995.7.1.108}.

\bibitem[Borsos et~al.(2023)Borsos, Marinier, Vincent, Kharitonov, Pietquin, Sharifi, Roblek, Teboul, Grangier, Tagliasacchi, et~al.]{borsos2023audiolm}
Zal{\'a}n Borsos, Rapha{\"e}l Marinier, Damien Vincent, Eugene Kharitonov, Olivier Pietquin, Matt Sharifi, Dominik Roblek, Olivier Teboul, David Grangier, Marco Tagliasacchi, et~al.
\newblock {AudioLM}: a language modeling approach to audio generation.
\newblock \emph{IEEE/ACM transactions on audio, speech, and language processing}, 31:\penalty0 2523--2533, 2023.

\bibitem[Camuto et~al.(2020)Camuto, Willetts, Simsekli, Roberts, and Holmes]{camuto2020explicit}
Alexander Camuto, Matthew Willetts, Umut Simsekli, Stephen~J Roberts, and Chris~C Holmes.
\newblock Explicit regularisation in {Gaussian} noise injections.
\newblock \emph{Advances in Neural Information Processing Systems}, 33:\penalty0 16603--16614, 2020.

\bibitem[Chan et~al.(2024)Chan, Corona, Park, Cho, Bai, and Darrell]{chan2024analyzing}
David~M Chan, Rodolfo Corona, Joonyong Park, Cheol~Jun Cho, Yutong Bai, and Trevor Darrell.
\newblock Analyzing the language of visual tokens.
\newblock \emph{arXiv preprint arXiv:2411.05001}, 2024.

\bibitem[Chen et~al.(2022)Chen, Wang, Chen, Wu, Liu, Chen, Li, Kanda, Yoshioka, Xiao, et~al.]{chen2022wavlm}
Sanyuan Chen, Chengyi Wang, Zhengyang Chen, Yu~Wu, Shujie Liu, Zhuo Chen, Jinyu Li, Naoyuki Kanda, Takuya Yoshioka, Xiong Xiao, et~al.
\newblock {WavLM}: Large-scale self-supervised pre-training for full stack speech processing.
\newblock \emph{IEEE Journal of Selected Topics in Signal Processing}, 16\penalty0 (6):\penalty0 1505--1518, 2022.

\bibitem[Chinen et~al.(2020)Chinen, Lim, Skoglund, Gureev, O'Gorman, and Hines]{chinen2020visqol}
Michael Chinen, Felicia~SC Lim, Jan Skoglund, Nikita Gureev, Feargus O'Gorman, and Andrew Hines.
\newblock {ViSQOL} v3: An open source production ready objective speech and audio metric.
\newblock In \emph{2020 twelfth international conference on quality of multimedia experience (QoMEX)}, pages 1--6. IEEE, 2020.

\bibitem[D{\'e}fossez et~al.(2022)D{\'e}fossez, Copet, Synnaeve, and Adi]{defossez2022high}
Alexandre D{\'e}fossez, Jade Copet, Gabriel Synnaeve, and Yossi Adi.
\newblock High fidelity neural audio compression.
\newblock \emph{arXiv preprint arXiv:2210.13438}, 2022.

\bibitem[D\'efossez et~al.(2024)D\'efossez, Mazar\'e, Orsini, Royer, P\'erez, J\'egou, Grave, and Zeghidour]{kyutai2024moshi}
Alexandre D\'efossez, Laurent Mazar\'e, Manu Orsini, Am\'elie Royer, Patrick P\'erez, Herv\'e J\'egou, Edouard Grave, and Neil Zeghidour.
\newblock Moshi: a speech-text foundation model for real-time dialogue.
\newblock Technical report, 2024.
\newblock URL \url{https://arxiv.org/abs/2410.00037}.

\bibitem[Devlin et~al.(2019)Devlin, Chang, Lee, and Toutanova]{devlin2019bert}
Jacob Devlin, Ming-Wei Chang, Kenton Lee, and Kristina Toutanova.
\newblock {BERT}: Pre-training of deep bidirectional transformers for language understanding.
\newblock In \emph{Proc. NAACL}, pages 4171--4186, 2019.

\bibitem[Du et~al.(2024)Du, Ai, Zheng, and Ling]{du2024apcodec+}
Hui-Peng Du, Yang Ai, Rui-Chen Zheng, and Zhen-Hua Ling.
\newblock {APCodec+}: A spectrum-coding-based high-fidelity and high-compression-rate neural audio codec with staged training paradigm.
\newblock In \emph{2024 IEEE 14th International Symposium on Chinese Spoken Language Processing (ISCSLP)}, pages 676--680. IEEE, 2024.

\bibitem[Gong et~al.(2022)Gong, Yu, and Glass]{gong2022vocalsound}
Yuan Gong, Jin Yu, and James Glass.
\newblock {VocalSound}: A dataset for improving human vocal sounds recognition.
\newblock In \emph{Proc. ICASSP}, pages 151--155. IEEE, 2022.

\bibitem[Huang et~al.(2022)Huang, Xu, Li, Baevski, Auli, Galuba, Metze, and Feichtenhofer]{huang2022masked}
Po-Yao Huang, Hu~Xu, Juncheng Li, Alexei Baevski, Michael Auli, Wojciech Galuba, Florian Metze, and Christoph Feichtenhofer.
\newblock Masked autoencoders that listen.
\newblock \emph{Advances in Neural Information Processing Systems}, 35:\penalty0 28708--28720, 2022.

\bibitem[Huang et~al.(2023)Huang, Meng, and Ko]{huang2023repcodec}
Zhichao Huang, Chutong Meng, and Tom Ko.
\newblock Repcodec: A speech representation codec for speech tokenization.
\newblock \emph{arXiv preprint arXiv:2309.00169}, 2023.

\bibitem[Ji et~al.(2024)Ji, Jiang, Wang, Chen, Fang, Zuo, Yang, Cheng, Wang, Li, et~al.]{ji2024wavtokenizer}
Shengpeng Ji, Ziyue Jiang, Wen Wang, Yifu Chen, Minghui Fang, Jialong Zuo, Qian Yang, Xize Cheng, Zehan Wang, Ruiqi Li, et~al.
\newblock {WavTokenizer}: an efficient acoustic discrete codec tokenizer for audio language modeling.
\newblock \emph{arXiv preprint arXiv:2408.16532}, 2024.

\bibitem[Kahn et~al.(2020)Kahn, Riviere, Zheng, Kharitonov, Xu, Mazar{\'e}, Karadayi, Liptchinsky, Collobert, Fuegen, et~al.]{kahn2020libri}
Jacob Kahn, Morgane Riviere, Weiyi Zheng, Evgeny Kharitonov, Qiantong Xu, Pierre-Emmanuel Mazar{\'e}, Julien Karadayi, Vitaliy Liptchinsky, Ronan Collobert, Christian Fuegen, et~al.
\newblock Libri-light: A benchmark for {ASR} with limited or no supervision.
\newblock In \emph{Proc. ICASSP}, pages 7669--7673. IEEE, 2020.

\bibitem[Kong et~al.(2020)Kong, Kim, and Bae]{kong2020hifi}
Jungil Kong, Jaehyeon Kim, and Jaekyoung Bae.
\newblock {HiFi-GAN}: Generative adversarial networks for efficient and high fidelity speech synthesis.
\newblock \emph{Advances in neural information processing systems}, 33:\penalty0 17022--17033, 2020.

\bibitem[Kumar et~al.(2023)Kumar, Seetharaman, Luebs, Kumar, and Kumar]{kumar2023high}
Rithesh Kumar, Prem Seetharaman, Alejandro Luebs, Ishaan Kumar, and Kundan Kumar.
\newblock High-fidelity audio compression with improved {RVQGAN}.
\newblock \emph{Advances in Neural Information Processing Systems}, 36:\penalty0 27980--27993, 2023.

\bibitem[Liu et~al.(2024{\natexlab{a}})Liu, Feng, Xue, Wang, Wu, Lu, Zhao, Deng, Zhang, Ruan, et~al.]{liu2024deepseek}
Aixin Liu, Bei Feng, Bing Xue, Bingxuan Wang, Bochao Wu, Chengda Lu, Chenggang Zhao, Chengqi Deng, Chenyu Zhang, Chong Ruan, et~al.
\newblock Deepseek-v3 technical report.
\newblock \emph{arXiv preprint arXiv:2412.19437}, 2024{\natexlab{a}}.

\bibitem[Liu et~al.(2024{\natexlab{b}})Liu, Xu, Yuan, Wu, Wang, and Plumbley]{liu2024semanticodec}
Haohe Liu, Xuenan Xu, Yi~Yuan, Mengyue Wu, Wenwu Wang, and Mark~D Plumbley.
\newblock {SemantiCodec}: An ultra low bitrate semantic audio codec for general sound.
\newblock \emph{IEEE Journal of Selected Topics in Signal Processing}, 2024{\natexlab{b}}.

\bibitem[Loshchilov and Hutter(2017)]{loshchilov2017decoupled}
Ilya Loshchilov and Frank Hutter.
\newblock Decoupled weight decay regularization.
\newblock \emph{arXiv preprint arXiv:1711.05101}, 2017.

\bibitem[Pan et~al.(2024)Pan, Ma, and Zhao]{pan2024promptcodec}
Yu~Pan, Lei Ma, and Jianjun Zhao.
\newblock {PromptCodec}: High-fidelity neural speech codec using disentangled representation learning based adaptive feature-aware prompt encoders.
\newblock \emph{arXiv e-prints}, pages arXiv--2404, 2024.

\bibitem[Panayotov et~al.(2015)Panayotov, Chen, Povey, and Khudanpur]{panayotov2015librispeech}
Vassil Panayotov, Guoguo Chen, Daniel Povey, and Sanjeev Khudanpur.
\newblock Librispeech: an {ASR} corpus based on public domain audio books.
\newblock In \emph{Proc. ICASSP}, pages 5206--5210. IEEE, 2015.

\bibitem[Piczak()]{piczak2015dataset}
Karol~J. Piczak.
\newblock {ESC}: {Dataset} for {Environmental Sound Classification}.
\newblock In \emph{Proceedings of the 23rd {Annual ACM Conference} on {Multimedia}}, pages 1015--1018. {ACM Press}.
\newblock ISBN 978-1-4503-3459-4.
\newblock \doi{10.1145/2733373.2806390}.
\newblock URL \url{http://dl.acm.org/citation.cfm?doid=2733373.2806390}.

\bibitem[Radford et~al.(2019)Radford, Wu, Child, Luan, Amodei, and Sutskever]{radford2019language}
Alec Radford, Jeff Wu, Rewon Child, David Luan, Dario Amodei, and Ilya Sutskever.
\newblock Language models are unsupervised multitask learners.
\newblock 2019.

\bibitem[Radford et~al.(2022)Radford, Kim, Xu, Brockman, McLeavey, and Sutskever]{radford2022whisper}
Alec Radford, Jong~Wook Kim, Tao Xu, Greg Brockman, Christine McLeavey, and Ilya Sutskever.
\newblock Robust speech recognition via large-scale weak supervision, 2022.
\newblock URL \url{https://arxiv.org/abs/2212.04356}.

\bibitem[Rahaman et~al.(2019)Rahaman, Baratin, Arpit, Draxler, Lin, Hamprecht, Bengio, and Courville]{rahaman2019spectral}
Nasim Rahaman, Aristide Baratin, Devansh Arpit, Felix Draxler, Min Lin, Fred Hamprecht, Yoshua Bengio, and Aaron Courville.
\newblock On the spectral bias of neural networks.
\newblock In \emph{International conference on machine learning}, pages 5301--5310. PMLR, 2019.

\bibitem[Rix et~al.(2001)Rix, Beerends, Hollier, and Hekstra]{PESQ}
A.W. Rix, J.G. Beerends, M.P. Hollier, and A.P. Hekstra.
\newblock Perceptual evaluation of speech quality ({PESQ})-a new method for speech quality assessment of telephone networks and codecs.
\newblock In \emph{Proc. ICASSP}, volume~2, pages 749--752 vol.2, 2001.
\newblock \doi{10.1109/ICASSP.2001.941023}.

\bibitem[Saeki et~al.(2022)Saeki, Xin, Nakata, Koriyama, Takamichi, and Saruwatari]{saeki2022utmos}
Takaaki Saeki, Detai Xin, Wataru Nakata, Tomoki Koriyama, Shinnosuke Takamichi, and Hiroshi Saruwatari.
\newblock {UTMOS}: Utokyo-sarulab system for {voiceMOS} challenge 2022.
\newblock \emph{arXiv preprint arXiv:2204.02152}, 2022.

\bibitem[Siuzdak(2023)]{siuzdak2023vocos}
Hubert Siuzdak.
\newblock Vocos: Closing the gap between time-domain and fourier-based neural vocoders for high-quality audio synthesis.
\newblock \emph{arXiv preprint arXiv:2306.00814}, 2023.

\bibitem[Siuzdak et~al.(2024)Siuzdak, Gr{\"o}tschla, and Lanzend{\"o}rfer]{siuzdak2024snac}
Hubert Siuzdak, Florian Gr{\"o}tschla, and Luca~A Lanzend{\"o}rfer.
\newblock {SNAC}: Multi-scale neural audio codec.
\newblock \emph{arXiv preprint arXiv:2410.14411}, 2024.

\bibitem[Skorokhodov et~al.(2025)Skorokhodov, Girish, Hu, Menapace, Li, Abdal, Tulyakov, and Siarohin]{skorokhodov2025improving}
Ivan Skorokhodov, Sharath Girish, Benran Hu, Willi Menapace, Yanyu Li, Rameen Abdal, Sergey Tulyakov, and Aliaksandr Siarohin.
\newblock Improving the diffusability of autoencoders.
\newblock \emph{arXiv preprint arXiv:2502.14831}, 2025.

\bibitem[Stanisz et~al.(2024)Stanisz, Dro{\.z}d{\.z}, and Kwapie{\'n}]{stanisz2024complex}
Tomasz Stanisz, Stanis{\l}aw Dro{\.z}d{\.z}, and Jaros{\l}aw Kwapie{\'n}.
\newblock Complex systems approach to natural language.
\newblock \emph{Physics Reports}, 1053:\penalty0 1--84, 2024.

\bibitem[Taal et~al.(2010)Taal, Hendriks, Heusdens, and Jensen]{STOI}
Cees~H. Taal, Richard~C. Hendriks, Richard Heusdens, and Jesper Jensen.
\newblock A short-time objective intelligibility measure for time-frequency weighted noisy speech.
\newblock In \emph{2010 IEEE International Conference on Acoustics, Speech and Signal Processing}, pages 4214--4217, 2010.
\newblock \doi{10.1109/ICASSP.2010.5495701}.

\bibitem[Van Den~Oord et~al.(2017)Van Den~Oord, Vinyals, et~al.]{van2017neural}
Aaron Van Den~Oord, Oriol Vinyals, et~al.
\newblock Neural discrete representation learning.
\newblock \emph{Advances in neural information processing systems}, 30, 2017.

\bibitem[Vaswani et~al.(2017)Vaswani, Shazeer, Parmar, Uszkoreit, Jones, Gomez, Kaiser, and Polosukhin]{vaswani2017attention}
Ashish Vaswani, Noam Shazeer, Niki Parmar, Jakob Uszkoreit, Llion Jones, Aidan~N Gomez, {\L}ukasz Kaiser, and Illia Polosukhin.
\newblock Attention is all you need.
\newblock \emph{Advances in neural information processing systems}, 30, 2017.

\bibitem[Wu et~al.(2024)Wu, Kanda, Eskimez, and Li]{wu2024ts3}
Haibin Wu, Naoyuki Kanda, Sefik~Emre Eskimez, and Jinyu Li.
\newblock Ts3-codec: Transformer-based simple streaming single codec.
\newblock \emph{arXiv preprint arXiv:2411.18803}, 2024.

\bibitem[Xin et~al.(2024)Xin, Tan, Takamichi, and Saruwatari]{xin2024bigcodec}
Detai Xin, Xu~Tan, Shinnosuke Takamichi, and Hiroshi Saruwatari.
\newblock {BigCodec}: Pushing the limits of low-bitrate neural speech codec.
\newblock \emph{arXiv preprint arXiv:2409.05377}, 2024.

\bibitem[Yang et~al.(2024)Yang, Yang, Zhang, Hui, Zheng, Yu, Li, Liu, Huang, Wei, et~al.]{yang2024qwen2}
An~Yang, Baosong Yang, Beichen Zhang, Binyuan Hui, Bo~Zheng, Bowen Yu, Chengyuan Li, Dayiheng Liu, Fei Huang, Haoran Wei, et~al.
\newblock Qwen2.5 technical report.
\newblock \emph{arXiv preprint arXiv:2412.15115}, 2024.

\bibitem[Yao et~al.(2025)Yao, Yang, and Wang]{yao2025reconstruction}
Jingfeng Yao, Bin Yang, and Xinggang Wang.
\newblock Reconstruction vs. generation: Taming optimization dilemma in latent diffusion models.
\newblock \emph{arXiv preprint arXiv:2501.01423}, 2025.

\bibitem[Ye et~al.(2025)Ye, Sun, Lei, et~al.]{ye2025codec}
Zhen Ye, Peiwen Sun, Jiahe Lei, et~al.
\newblock Codec does matter: Exploring the semantic shortcoming of codec for audio language model.
\newblock In \emph{Proceedings of the AAAI Conference on Artificial Intelligence}, 2025.

\bibitem[Zeghidour et~al.(2021)Zeghidour, Luebs, Omran, Skoglund, and Tagliasacchi]{zeghidour2021soundstream}
Neil Zeghidour, Alejandro Luebs, Ahmed Omran, Jan Skoglund, and Marco Tagliasacchi.
\newblock {SoundStream}: An end-to-end neural audio codec.
\newblock \emph{IEEE/ACM Transactions on Audio, Speech, and Language Processing}, 30:\penalty0 495--507, 2021.

\bibitem[Zhang et~al.(2023)Zhang, Zhang, Li, Zhou, and Qiu]{zhang2023speechtokenizer}
Xin Zhang, Dong Zhang, Shimin Li, Yaqian Zhou, and Xipeng Qiu.
\newblock {SpeechTokenizer}: Unified speech tokenizer for speech large language models.
\newblock \emph{arXiv preprint arXiv:2308.16692}, 2023.

\bibitem[Zhao et~al.(2024)Zhao, Zou, Shah, and Liu]{zhao2024representation}
Wenhao Zhao, Qiran Zou, Rushi Shah, and Dianbo Liu.
\newblock Representation collapsing problems in vector quantization.
\newblock \emph{arXiv preprint arXiv:2411.16550}, 2024.

\bibitem[Zheng et~al.(2024)Zheng, Tu, Kang, Chen, Zhang, Xiao, Yang, and Ma]{zheng2024freecodec}
Youqiang Zheng, Weiping Tu, Yueteng Kang, Jie Chen, Yike Zhang, Li~Xiao, Yuhong Yang, and Long Ma.
\newblock {FreeCodec}: A disentangled neural speech codec with fewer tokens.
\newblock \emph{arXiv preprint arXiv:2412.01053}, 2024.

\bibitem[Zhou et~al.(2022)Zhou, Sisman, Liu, and Li]{zhou2022emotional}
Kun Zhou, Berrak Sisman, Rui Liu, and Haizhou Li.
\newblock Emotional voice conversion: Theory, databases and esd.
\newblock \emph{Speech Communication}, 137:\penalty0 1--18, 2022.

\bibitem[Zhu et~al.(2024{\natexlab{a}})Zhu, Wei, Lu, and Chen]{zhu2024scaling}
Lei Zhu, Fangyun Wei, Yanye Lu, and Dong Chen.
\newblock Scaling the codebook size of vqgan to 100,000 with a utilization rate of 99\%.
\newblock \emph{arXiv preprint arXiv:2406.11837}, 2024{\natexlab{a}}.

\bibitem[Zhu et~al.(2024{\natexlab{b}})Zhu, Li, Xin, and Xu]{zhu2024addressing}
Yongxin Zhu, Bocheng Li, Yifei Xin, and Linli Xu.
\newblock Addressing representation collapse in vector quantized models with one linear layer.
\newblock \emph{arXiv preprint arXiv:2411.02038}, 2024{\natexlab{b}}.

\end{thebibliography}
\bibliographystyle{plainnat}
\newpage
\appendix

\section{Technical Proofs}

\subsection{\texorpdfstring{Proof of Proposition \hyperref[prop1]{1}}{Proof of Proposition 1}}

\label{appendix:prop1}

Let \(\boldsymbol{\epsilon}\) be an isotropic random Gaussian vector in \(\mathbb{R}^d\) with
$\boldsymbol{\epsilon}\sim\mathcal{N}(\mathbf{0},\sigma^2 I)$, 
its probability density function \(p_{\boldsymbol{\epsilon}}\colon\mathbb{R}^d\to\mathbb{R}\) is
\[
p_{\boldsymbol{\epsilon}}(\mathbf{u})
=
\frac{1}{(2\pi\sigma^2)^{d/2}}
e^{-\frac{\|\mathbf{u}\|^2}{2\sigma^2}}
\]

Denote by $P_{\boldsymbol{\epsilon}}$ the corresponding Gaussian probability measure on $\mathbb{R}^d$, then for any Borel set $A\subset\mathbb{R}^d$, we have 
$
P_{\boldsymbol{\epsilon}}(A)
= \int_{A} p_{\boldsymbol{\epsilon}}(\mathbf{u})\,d\mathbf{u}
$ , where $\mathbf{u}$ denotes the Lebesgue measure on \(\mathbb{R}^d\).

Then for any measurable function \(f\), we have 
\[
\mathbb{E}\bigl[f(\mathbf{x}+\boldsymbol{\epsilon})\bigr]
= \int_{\mathbb{R}^d} f(\mathbf{x}+\mathbf{u})\,dP_{\boldsymbol{\epsilon}}(\mathbf{u})
= \bigl(p_{\boldsymbol{\epsilon}} * f\bigr)(\mathbf{x})
\]

Let \(\mathbf{m}\sim\mathrm{Bernoulli}(p)^d\), \(\mathbf{m}\in\{0,1\}^d\), and Gaussian noise \(\boldsymbol{\epsilon}\sim\mathcal{N}(\mathbf{0},\sigma^2 I)\). Let the neural network be \(f\) with input \(\mathbf{x} \in \mathbb{R}^d\). 
And $\mathbf{m}$, $\boldsymbol{\epsilon}$, and $\mathbf{x}$ are pairwise independent. 
Each frame of $\mathbf{x}$ is replaced with the independent Gaussian noise with probability $p$. Then we have 
\[
\tilde{x}_i =
\begin{cases}
x_i, & m_i = 0,\\
\epsilon_i, & m_i = 1,
\end{cases}
\quad i = 1,\dots,d.
\]
Equivalently, the input $\tilde{\mathbf{x}}$ satisfies
\[
\tilde{\mathbf{x}} = (\mathbf{1} - \mathbf{m}) \odot \mathbf{x} \;+\; \mathbf{m} \odot \boldsymbol{\epsilon}.
\]
where $\odot$ means the Hadamard product. We can write $\tilde{\mathbf{x}}$ in  additive form $\tilde{\mathbf{x}} = \mathbf{x} + \boldsymbol{\epsilon}'_{p,\sigma}$, where $\boldsymbol{\epsilon}'_{p,\sigma} = \mathbf{m} \odot (\boldsymbol{\epsilon} - \mathbf{x})$. 

\begin{align*}
\mathbb{E}_{\boldsymbol{\epsilon}'_{p,\sigma}} \bigl[ f(\mathbf{x} + \boldsymbol{\epsilon}'_{p,\sigma}) \bigr]
&= \int_{\mathbb{R}^d} f(\mathbf{x} + \mathbf{u})\, \mathrm{d}P_{\boldsymbol{\epsilon}'_{p,\sigma}}(\mathbf{u}) \\
&= (1 - p)\, f(\mathbf{x}) + p \cdot (k_\sigma * f)(\mathbf{x}) \\
&= (g_{p,\sigma} * f)(\mathbf{x})
\end{align*}
where $k_\sigma(\mathbf{x})
= (2\pi \sigma^2)^{\tfrac{d}{2}}
\exp\bigl(- \|\mathbf{x}\|^2 / (2\sigma^2)\bigr)$ is the Gaussian kernel and $g_{p,\sigma}$ is defined by 
$
g_{p,\sigma}(\mathbf{x})
= (1 - p)\delta(\mathbf{x})
+ p k_{\sigma}(\mathbf{x})
$.

Taking the Fourier transform (denoted by $\widehat{\cdot}$ ), and using $\widehat{\delta}( \boldsymbol{\omega} ) = 1$ and 
$\widehat{k}_{\sigma}( \boldsymbol{\omega} ) = e^{-\sigma^{2} \| \boldsymbol{\omega} 
 \|^{2}/2}$, we obtain
\[
\widehat{g}_{p,\sigma}(\boldsymbol{\omega})
= (1 - p) + p\,e^{-\sigma^{2}\|\boldsymbol{\omega}\|^{2}/2}
\tag{2}
\]

Thus, for any Fourier‐transformable function $f$, we have 
\[
\mathbb{E}\bigl[f(\tilde{\mathbf{x}})\bigr]
= \bigl[(1 - p) + p\,e^{-\sigma^{2}\|\boldsymbol{\omega}\|^{2}/2}\bigr]\;\widehat{f}(\boldsymbol{\omega})\,.
\]

\section{Evaluation Metrics}
\label{appendix:metrics}

Below we define the evaluation metrics used in this paper. Metrics annotated with ↑ indicate that higher values are better, whereas those annotated with ↓ imply that lower values are preferable.

\paragraph{Computation Efficiency}
\begin{itemize}
    \item Model Parameter Count (nParams): The total number of model parameters.
    \item Bitrate: The number of bits transmitted or stored per second.
    \item Frame Rate: The number of frames the encoder processes or outputs per second.
    \item Token Rate: The rate at which discrete tokens are emitted by the quantizer.
    \item Streaming Capability: A boolean flag indicating whether encoding/decoding can be performed on a per-frame, online basis. \textit{True} enables low-latency, real-time processing. \textit{False} requires batch inputs and incurs higher end-to-end delay.
    \item Number of Codebook Layers: The number of layers in a multi-layer or hierarchical quantization scheme. Increasing the number of layers enhances quantization expressiveness but also raises storage and lookup overhead, and imposes additional overhead in downstream task modeling.
\end{itemize}

\paragraph{Speech Intelligibility}
\begin{itemize}
    \item Short-Time Objective Intelligibility (STOI ↑): An objective measure of speech intelligibility based on short-time signal alignment and correlation, ranging from 0 to 1~\maybeCite{STOI}. Higher STOI values indicate speech that is more easily understood by listeners.
    \item Word Error Rate (WER ↓): We utilize the official weights of the Whisper-large-v3~\maybeCite{radford2022whisper} model to transcribe the synthesized audio and compute the Word Error Rate (WER), a metric indicative of the audio’s clarity.
    \item Phone Error Rate (PER ↓): PER offers a finer-grained assessment of recognition performance compared to WER.
\end{itemize}

\paragraph{Distortion and Perceptual}
\begin{itemize}
    \item Perceptual Evaluation of Speech Quality (PESQ ↑): A standard that simulates subjective listening quality, with scores typically ranging from 1 to 5 \maybeCite{PESQ}.
    \item Virtual Speech Quality Objective Listener (ViSQOL ↑): An objective, full-reference metric for perceived audio quality. The scores range from 1 (the worst) to 5 (the best) \maybeCite{chinen2020visqol}.
    \item UTokyo-SaruLab MOS (UTMOS ↑): A machine-learning–based predictor of human MOS (Mean Opinion Score), generally in the range 1–5~\maybeCite{saeki2022utmos}.
\end{itemize}

\paragraph{Speaker Similarity}
\begin{itemize}
    \item Speaker Similarity (SPK-SIM ↑): We use this metric to quantify the consistency between decoded audio and the characteristics of the original speaker. We use the WavLM-based\footnote{\url{https://github.com/microsoft/UniSpeech/tree/main/downstreams/speaker_verification}} model ~\maybeCite{chen2022wavlm} for speaker verification.
\end{itemize}

\end{document}